\documentclass[a4paper,fleqn,usenatbib]{mnras}

\usepackage[T1]{fontenc}
\usepackage{ae,aecompl}
\usepackage{graphicx}
\usepackage{amsmath}
\usepackage{amssymb}
\usepackage{paralist}
\usepackage{xspace}
\usepackage{newtxtext,newtxmath}

\newcommand{\rxte}{\textsl{RXTE}\xspace}
\newcommand{\gx}{GX~304$-$1\xspace}
\newcommand{\msol}{\mathrm{M_\odot}}
\newcommand{\rsol}{\mathrm{R_\odot}}

\defcitealias{rothschild2017a}{R17}
\defcitealias{kopal1959a}{K59}
\defcitealias{alexander1976a}{A76}

\title{A precessing Be disk as a possible model for occultation events in \gx}

\author[M.~K\"uhnel et al.]
{M.~K\"uhnel,$^{1}$\thanks{E-mail: matthias.kuehnel@sternwarte.uni-erlangen.de}
R.~E.~Rothschild,$^{2}$
A.~T.~Okazaki,$^{3}$
S.~M\"uller,$^{1}$
K.~Pottschmidt,$^{4,5}$
\newauthor
R.~Ballhausen,$^{1}$
J.~Choi,$^{6}$
I.~Kreykenbohm,$^{1}$
F.~F\"urst,$^{7}$
D.M.~Marcu-Cheatham,$^{4,5}$
\newauthor
P.~Hemphill,$^{8}$
M.~Sagredo,$^{1}$
P.~Kretschmar,$^{7}$
S.~Mart\'inez-N\'u\~nez,$^{9}$
J.~M. Torrej\'on,$^{10}$
\newauthor
R.~Staubert,$^{11}$
and J.~Wilms$^{1}$
\\
$^{1}$Dr.\ Karl Remeis-Observatory \& ECAP, Universit\"at Erlangen-N\"urnberg,
Sternwartstr.~7, 96049 Bamberg, Germany\\
$^{2}$Center for Astrophysics and Space Sciences, University of California, San
Diego, La Jolla, CA 92093, USA\\
$^{3}$Faculty of Engineering, Hokkai-Gakuen University, Toyohira-ku, Sapporo 062-8605, Japan\\
$^{4}$CRESST/CSST/Department of Physics, UMBC, Baltimore, MD 21250, USA\\
$^{5}$NASA Goddard Space Flight Center, Greenbelt, MD 20771, USA\\
$^{6}$Harvard Smithsonian Center for Astrophysics, Cambridge, MA 02138, USA\\
$^{7}$European Space Astronomy Centre (ESA/ESAC), Operations Departement, 28691 Villanueva
de la Ca{\~n}ada, Madrid, Spain\\
$^{8}$MIT Kavli Institute for Astrophysics and Space Research, 77 Massachusetts Ave,
Cambridge, MA 02139, USA\\
$^{9}$Instituto de F\'isica de Cantabria, Avda.\ los Castros s/n, 39005
Santander, Spain\\
$^{10}$Instituto Universitario de F\'isica Aplicada a las Ciencias y las
Tecnolog\'ias, University of Alicante, P.O.\ Box 99, 03690 Alicante, Spain\\
$^{11}$Institut f\"ur Astronomie und Astrophysik, Universit\"at T\"ubingen,
Sand 1, 72076 T\"ubingen, Germany}

\date{Accepted XXX. Received YYY; in original form ZZZ}
\pubyear{2017}

\begin{document}
\maketitle

\begin{abstract}
We report on the \rxte detection of a sudden increase in the absorption
column density, $N_\mathrm{H}$, during the 2011 May outburst of \gx. The
$N_\mathrm{H}$ increased up to ${\sim}16\times
10^{22}$\,atoms\,cm$^{-2}$, which is a factor of 3--4 larger than what
is usually measured during the outbursts of \gx as covered by \rxte.
Additionally, an increase in the variability of the hardness ratio as
calculated from the energy resolved \rxte-PCA light curves is measured
during this time range. We interpret these facts as an occultation event
of the neutron star by material in the line of sight. Using a simple 3D
model of an inclined and precessing Be disk around the Be type
companion, we are able to qualitatively explain the $N_\mathrm{H}$
evolution over time. We are able to constrain the Be-disk density to be
on the order of $10^{-11}$\,g\,cm$^{-3}$. Our model strengthens the idea
of inclined Be disks as origin of double-peaked outbursts as the derived
geometry allows accretion twice per orbit under certain conditions.
\end{abstract}

\begin{keywords}
X-rays: binaries -- stars: Be -- stars: neutron -- occultations -- pulsars:
individual: \gx
\end{keywords}

\section{Introduction}

The H$\alpha$ emission line observed in the optical spectra of many
B-stars is believed to originate from a circumstellar disk \citep[see,
e.g.,][]{hanuschik1996a}. The H$\alpha$ line profile in these stars
often shows a double-peaked structure due to intrinsic rotation of the
disk. These emission features are denoted with an ``e'' in the stellar
spectral type, which is why these stars are known as Be stars. The
Be-star disks are supposed to form by the viscous diffusion of gas
ejected from the central star \citep{lee1991a}. They are geometrically
thin and rotating at Keplerian velocities \citep[see][for a recent
review]{rivinius2013a}.

Circumstellar disks around main-sequence stars have been proposed to
explain the transient nature of some X-ray binaries
\citep{rappaport1978a}. In these so-called Be X-ray binaries (BeXRBs)
a compact object is on a wide and eccentric orbit around a Be-type
companion star. Around the periastron passage mass transfer from the
Be-star disk becomes possible, which results in a sudden and luminous
X-ray outburst. Furthermore, the Be disk in BeXRBs is tidally
truncated at a certain resonance radius due to the presence of the
orbiting neutron star \citep{reig1997a,negueruela2001a,okazaki2002a}.
This truncation leads to higher surface densities of BeXRB-disks
compared to disks around isolated stars \citep[][in agreement with
findings by \citealt{zamanov2001a}]{okazaki2002a}. However,
\citet{zamanov2001a} have drawn this conclusion based on measurements
of H$\alpha$ equivalent widths and the peak separation of the line
profile.

BeXRBs offer the opportunity to measure directly the Be-disk density
independently of other energy ranges such as optical wavelengths. The
particle density in the line of sight determines the absorption of
X-rays at energies below $\sim$10\,keV. As the X-ray line of sight is
fixed to the neutron star, which moves along its orbit, the particle
density around the Be star can be probed.

The BeXRB \gx, detected in 1967 by a balloon experiment
\citep{lewin1968a}, consists of a shell Be-star of type B2Vne
\citep{mason1978a,parkes1980a} or B0.7Ve \citep{liu2006a} and a
neutron star which shows X-ray pulsations around $\sim$272\,s
\citep{mcclintock1977a}. It is located at a distance of 2.4(5)\,kpc
\citep{parkes1980a} probably behind the Coalsack Nebula, a dark
molecular cloud on the Southern sky at a distance of ${\sim}150$\,pc
\citep[see][and references therein]{nyman2008a}.

\gx often remained in states of quiescence as, e.g., between 1980 and
2008. On the other hand, it showed regular outbursts previously with
intervals of 132.5(4)\,d as found by \citet{priedhorsky1983a}, who
interpreted this periodicity as the orbital period of the system. This
makes \gx a prime example of sources with a highly unpredictable
outburst behaviour \citep{manousakis2008a,klochkov2012a}. The source
again underwent regular outbursts in late 2009 until mid
2013. Recently, \citet{sugizaki2015b} confirmed the orbital period to
be $P_\mathrm{orb} = 132.189(22)$\,d based on the outburst spacing in
the MAXI-GSC light curves during this series. Together with pulse
frequency measurements by \textsl{Fermi}-GBM \citep{finger2009a} they
derived the remaining orbital parameters of \gx (see
Table~\ref{tab:orbit}).

In rare cases BeXRBs undergo so-called double-peaked outbursts, where
two consecutive outbursts can be observed, which are not separated by
the usually observed phase of quiescence. In order to explain the
observed double-peaked outbursts of \gx in 2012, \citet{postnov2015a}
proposed that the Be disk is probably inclined with respect to the
orbital plane. This is in line with smoothed particle hydrodynamic (SPH)
simulations performed by \citet{okazaki2013a}, who concluded that the
occurrence of so-called ``giant'' type II X-ray outbursts are probably
triggered by a misaligned Be disk.

\begin{table}
  \caption{Orbital parameters of \gx after \citet[$\gamma$ free
  model]{sugizaki2015b}.}
  \centering
  \begin{tabular}{ll}
  \hline
  Orbital period, $P_\mathrm{orb}$ (d) & $= 132.189 \pm 0.022$ \\
  Projected semi-major axis, $a \sin i$ (lt-s) & $= 601 \pm 38$ \\
  Eccentricity, $e$ & $= 0.462\pm 0.019$ \\
  Time of periastron passage, $\tau$ (MJD) & $= 55425.6 \pm 0.5$ \\
  Longitude of periastron, $\omega$ ($^\circ$) & = $130.0 \pm 4.4$ \\
  \hline
  \end{tabular}
  \label{tab:orbit}
\end{table}

This paper is based on the spectral results of the accompanying paper
by \citet[hereafter R17]{rothschild2017a}, who analysed 72 \rxte
observations taken during four outbursts between March~2010 and
May~2011. Here, we particularly focus on the evolution of the
absorption column density, $N_\mathrm{H}$, over time. The paper is
organized as follows: in Section~\ref{sec:obs} we briefly describe the
data reduction process performed by \citetalias{rothschild2017a}.
Section~\ref{sec:data} gives a summary and discussion of their
spectral results, which are then modelled by a simple 3D-model of a
precessing Be disk in Sect.~\ref{sec:model}. We present our
conclusions in Sect.~\ref{sec:conclude}.

\section{Observations and Data Reduction}
\label{sec:obs}

\subsection{Observations}

\begin{figure}
  \includegraphics[width=\columnwidth]{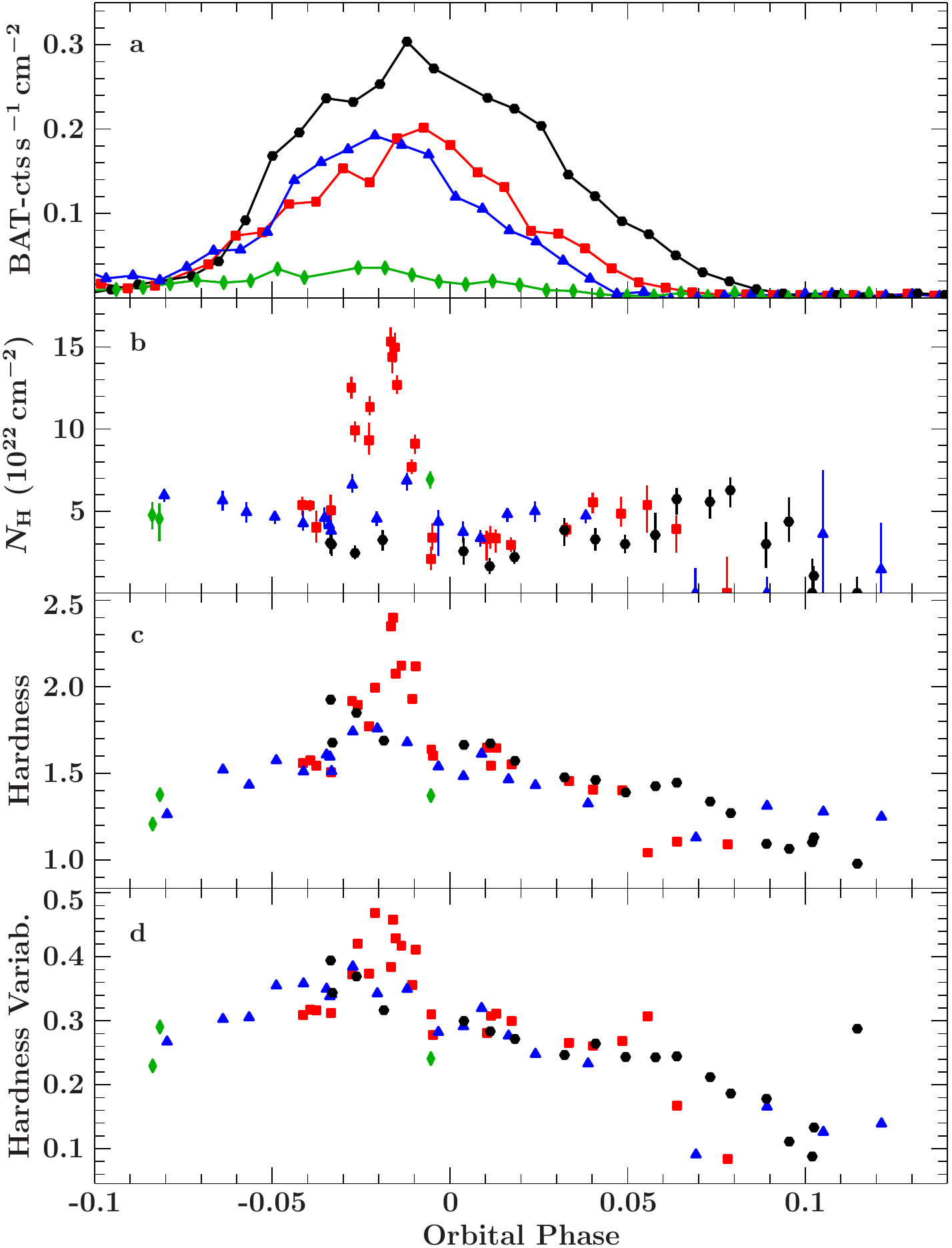}
  \caption{\textbf{a)} Count rate of \gx in \textsl{Swift}-BAT as a
    function of orbital phase during the outbursts in 2010~March/April
    (green diamonds), 2010~August (black circles), 2010~December (blue
    triangles), and 2011~May (red squares). \textbf{b)} Evolution of the
    absorption column density, $N_\mathrm{H}$, and \textbf{c)}
    evolution of the hardness ratio and \textbf{d)} its variability
    over the orbital phase.}
  \label{fig:parsphi}
\end{figure}

\rxte has observed \gx during four outbursts in 2010 and 2011 (see
Fig.~\ref{fig:parsphi}a). In total, 72 individual observations are
available with exposure times from a few 100\,s up to 18\,ks each.
The corresponding observation IDs and further basic informations
about each observation can be found in
\citetalias[Tab.~1]{rothschild2017a}. The exposure times of three
observations were less than the \gx's pulse period of $\sim$272\,s
\citep{mcclintock1977a}. These observations have been excluded from the
spectral analysis, which focuses on pulse phase averaged spectra.

\subsection{Data Reduction}

The data reduction process and the spectral results analysed further
in Section~\ref{sec:data} are the same as in \citetalias{rothschild2017a}.
In the following a brief summary of the data reduction is given.

The Proportional Counter Array \citep[PCA;][]{jahoda2006a} onboard \rxte
consisted of five identical Proportional Counter Units (PCU0--5). For
spectral analysis only data from the top layer of PCU2 have been
extracted since this PCU is known to be the best calibrated one
\citep{jahoda2006a}. We used the tools distributed by \texttt{HEASOFT}
v6.18 in order to extract spectra and lightcurves using the standard2f
data mode. We avoided PCU2-events up to 30 minutes from the start of the
previous passage of the South Atlantic Anomaly and for elevation angles
smaller than $10^\circ$ above the Earth's limb. Due to the high count
rate of \gx during many of the 72 observations we used 0.5 as the
highest accepted value for the electron ratio. For spectral analysis the
energy range of 3--60\,keV range has been used and no further spectral
binning was applied. Due to the high quality of the PCA-spectra during a
few observations, we had to add systematic uncertainties of 0.5\% at
channel energies $\le$15\,keV and of 1.5\% at higher energies in order
to achieve a reduced $\chi^2$, $\chi^2_\mathrm{red}$, near unity (see
\citetalias{rothschild2017a} for the affected ObsIDs). The PCU2 light
curves were extracted with a 16\,s time resolution in the detector
channels 4--15 and 15--60 (corresponding to the energy bands
2.9--7.7\,keV and 7.7--30.0\,keV, respectively). These channels have
been chosen such that the ratio of the light curves, i.e., a hardness
ratio, provides a handle on X-ray absorption (see
Sect.~\ref{sec:lcanalysis} for further details).

The High Energy X-ray Timing Experiment
\citep[HEXTE;][]{rothschild1998a} is sensitive for X-ray photons
between 15 and 250\,keV and consisted of two identical clusters,
HEXTE-A and -B. These clusters alternated between the on-source
position and two background positions. This so-called ``rocking''
mechanism allowed one to simultaneously measure the X-ray source and
background. Due to mechanical failures of this technique late in the
mission, cluster~A was fixed in the on-source position during all
observations of \gx, while cluster~B was fixed in one background
position 1.5$^\circ$ off-source. In order to investigate background
corrected spectra for \gx, we have extracted the source spectrum from
cluster~A and estimated its corresponding background from the
background spectrum of cluster~B using the \texttt{hextebackest} tool
as described in \citet{pottschmidt2006a} and
\citetalias[Appendix~A]{rothschild2017a}.  The resulting HEXTE-spectra
of \gx have been analysed in the 20--100\,keV energy range, no
spectral binning was applied, and no systematic uncertainties have
been added to the data.

\section{Data Analysis}
\label{sec:data}

\subsection{Absorption column density}

We have used the \textit{Interactive Spectral Interpretation System}
\citep[ISIS][]{houck2000a} v1.6.2-30 to perform a combined spectral
analysis of the PCA- and HEXTE-spectra for each \rxte observation. We
use the results of \citetalias{rothschild2017a}, who describe the
\rxte spectra with two different continuum models. The
\texttt{cutoffpl} model consists of a power-law in combination with a
multiplicative exponential and an additional blackbody
(\texttt{CUTOFFPL} $+$ \texttt{BBODY}). The second model
\texttt{highecut} is a power-law continuum with an exponential
roll-over, which sets in at higher photon energies (\texttt{POWERLAW}
$\times$ \texttt{HIGHECUT}). Both models take low energy X-ray
absorption into account using the model
\texttt{TBnew}\footnote{\url{http://pulsar.sternwarte.uni-erlangen.de/wilms/research/tbabs/}}
which is and extended version of the model by \citet{wilms2000a} with
interstellar element abundances. Here, the corresponding
cross-sections were set to those of \citet{verner1996b}. The emission
lines of iron at 6.40\,keV and 7.06\,keV were modelled with Gaussians
(\texttt{GAUSS}), and the cyclotron line known to be present in \gx
around 50--55\,keV \citep{yamamoto2011a} was described by a
multiplicative Gaussian absorption component (\texttt{GAUABS}).

Furthermore, as discussed by \citetalias{rothschild2017a} several
systematic features are detectable in the \rxte spectra. Although the
background for HEXTE-A can be estimated from HEXTE-B, the true
background still differs from the estimated one. This results in four
additional background lines at 30.17, 39.04, 53.0, and 66.64 keV,
which we have modelled by additional Gaussian components. In some
observations slight residuals in absorption are seen in PCA below
${\sim}$4.5\,keV, which we attribute to an insufficient modelling of
the Xenon L-edge and have modelled by a negative Gaussian fixed at
3.88\,keV. A similar systematic feature appears at 10.5\,keV for very
high source fluxes, which we included in the model as well. Finally,
we have introduced a flux calibration constant for HEXTE with respect
to PCA and allowed the fit to adapt the background strength for both
detectors (\texttt{CORBACK} in ISIS, which is similar to \texttt{RECOR}
in XSPEC). A detailed discussion of these systematic features in the
\rxte spectra, in particular the HEXTE background lines, is given in
\citetalias[Appendix~C]{rothschild2017a}.

The resulting evolution of the spectral parameters, especially those of
the power-law continuum and cyclotron line, for both continuum models
(\texttt{cutoffpl} and \texttt{highecut}, see above) is presented in
\citetalias{rothschild2017a}. In this work, we focus on the time
evolution of the absorption column density, $N_\mathrm{H}$, which tracks
the number of particles along the line of sight to the neutron star. As
already noticed by \citetalias{rothschild2017a} a large enhancement of
the column density is detected during the 2011~May outburst (see
Fig.~\ref{fig:parsphi}b), where $N_\mathrm{H}$ is about ${\times}3$
higher than what is usually observed in \gx with \rxte. This enhancement
event lasts for around 3~days until $N_\mathrm{H}$ suddenly drops to the
usual value. A weaker event (${\times}1.5$) is also visible during the
2010~December outburst. This result is independent of the choice of
continuum model (\texttt{cutoffpl} and \texttt{highecut}) and, thus, it
is unlikely that the low energy continuum of the neutron star has been
modelled improperly during these enhancement events. We will restrict
the following discussions to the results of the \texttt{cutoffpl} model.

\subsection{Light curve analysis}\label{sec:lcanalysis}

In order to confirm the enhancement in $N_\mathrm{H}$ in an
independent way of any phenomenological modelling of the X-ray
spectra, we have investigated the energy resolved \rxte light curves
of \gx. Since material in the line of sight results in absorption of
X-ray photons at energies ${\lesssim}10$\,keV, the ratio between two
light curves in different energy bands, a so-called hardness ratio,
then tracks absorption variability without analyzing the full X-ray
spectrum. We have computed the ratio of two PCA light curves (1\,s time
resolution) in the energy bands 2.9--7.7\,keV and 7.7--30.0\,keV. Due to
the energy dependence of \gx's pulse profile \citep[see,
e.g.,][]{jaisawal2016a}, residual pulsations are still visible in the
resulting hardness ratio over time, but the pulsed fraction is lower by
a factor of 3--5 than compared to the individual light curves.

Figure~\ref{fig:parsphi}c shows the resulting mean hardness ratio for
each observation as determined from the energy resolved light curves.
The hardness as defined via the energy bands given above is between
1.0 and 1.8 during most observations. At the same time when the
enhancement in the absorption column density is observed during the
2011\,May outburst (Fig.~\ref{fig:parsphi}b), the hardness suddenly
increases up to ${\sim}2.5$, i.e., by a factor of ${\sim}1.4$. Outside
of the enhancement event the hardness is again consistent with the data
from the other outbursts. This confirms the results of our spectral
analysis.

We also found that the standard deviation of the hardness ratio over
time during each observation (Figure~\ref{fig:parsphi}d) shows a very
similar increase by a factor of ${\sim}1.5$ during the 2011\,May
outburst. We interpret this variability as evidence for inhomogeneities
within the absorbing material in the line of sight, such as a clumpy
wind \citep[see, e.g.,][]{hemphill2014a}.

\section{Possible occultation by the Be disk}
\label{sec:model}

In order to investigate the origin of these enhancement events, we have
used the orbital parameters by \citet[see Table~\ref{tab:orbit}]{sugizaki2015b}
to convert the date of each \rxte observation into an orbital phase.
Figure~\ref{fig:parsphi}b shows the measured absorption column density
during all four outbursts over the resulting orbital phases. The two
enhancement events nicely align between the orbital phases $-0.03$ and
$-0.01$, which is right before the periastron passage of the neutron
star. In these events no significant flare was detected on top of the
\textsl{Swift}-BAT light curve of \gx (compare Fig.~\ref{fig:parsphi}a),
which is in agreement with the \rxte light curves \citepalias[see Fig.~1
of][]{rothschild2017a}. Thus, it is unlikely that the matter, which is
responsible for the enhancement in $N_\mathrm{H}$, was accreted by the
neutron star as this would result in an additional increase in X-ray
luminosity.

From \gx's orbital parameters we can estimate the dimension of the cloud
of material within the line of sight. Assuming that the material is
stationary and neglecting the eccentricity of the neutron star's orbit,
we find $2 \pi (a \sin i) \Delta t / P_\mathrm{orb} \sim 86\,$lt-s as
the dimension of the cloud with the duration $\Delta t{=}3$\,d of the
enhancement event in 2011~May. Such small scale density fluctuations are
not expected in a molecular cloud such as the Coalsack Nebula. Thus, the
cloud of material has to be located within the \gx system. The
enhancement event took place right before the periastron passage of the
neutron star. The location of the periastron within the binary system is
determined by its longitude, $\omega = 130^\circ$ (see
Table~\ref{tab:orbit}), which means that this point, as seen from Earth,
is behind the companion's tangent plane of the sky. That is, the neutron
star is farther away from Earth during the periastron passage than the
Be companion. Thus, the X-ray line of sight, which is always fixed on
the neutron star, might has passed through or behind the circumstellar
material of the companion, such as its equatorial disk.

In this Section, we investigate the possibility of an occultation
event of the neutron star by the Be disk. Therefore we introduce a
simple 3D-model of a rigid Be disk with a physically motivated
density profile and apply this model to the observed $N_\mathrm{H}$
evolution.

\subsection{Density profile of the Be disk}\label{sec:diskprofile}

\begin{figure}
  \includegraphics[width=\columnwidth]{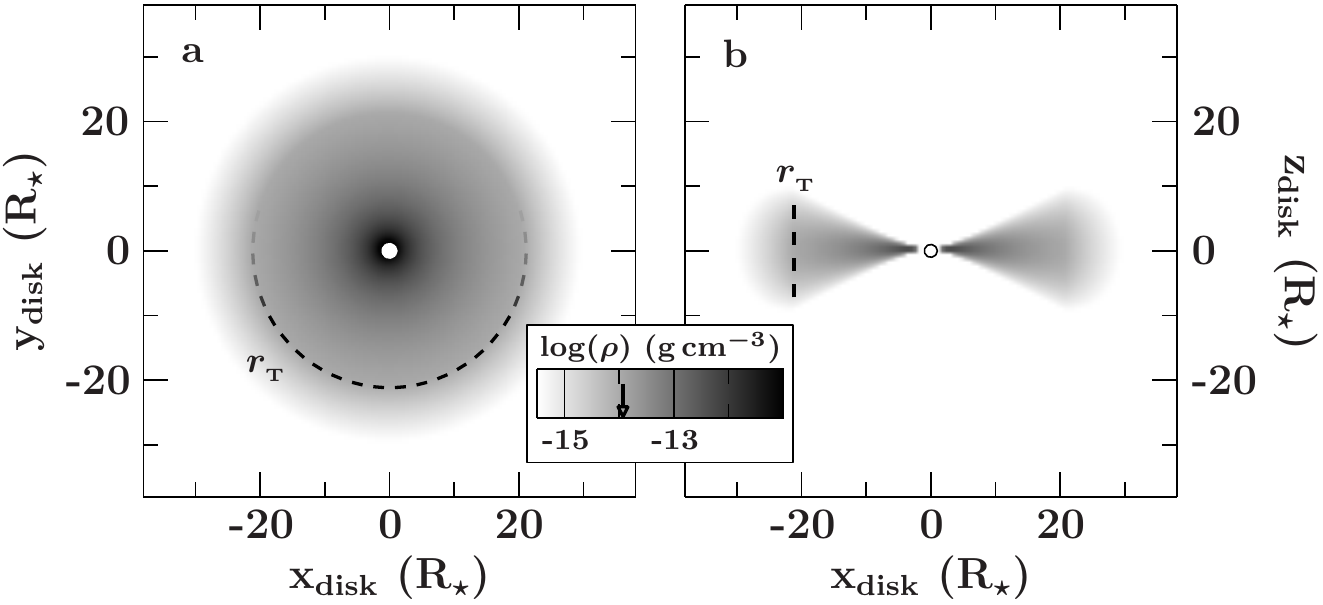}
  \caption{Assumed density profile of the Be disk as calculated
  after Eqs.~\mbox{\ref{eq:diskrho}--\ref{eq:diskrhotrunc}} for a
  base density of $\rho_0 = 10^{-11}$\,g\,cm$^{-3}$, a truncation
  radius $r_{_\mathrm{T}}/a = 0.49$, and a semi-major axis of $a=601$\,lt-s
  in Cartesian coordinates ($x_\mathrm{disk}$, $y_\mathrm{disk}$,
  $z_\mathrm{disk}$). The white circle in the disk's centre is the
  Be star. The arrow in the density scale marks the density at the
  truncation radius. \textbf{a)} Density profile within the disk
  plane ($z_\mathrm{disk}=0$) showing the radial dependence.
  \textbf{b)} Height dependence of the density profile, i.e.,
  perpendicular to the disk plane ($y_\mathrm{disk}=0$).}
  \label{fig:disksetup}
\end{figure}

According to the viscous decretion disk scenario \citep{lee1991a},
Be disks are formed by viscous diffusion of gas ejected from the
central star. They are Keplerian disks, radially supported by the
rotation, and in hydrostatic equilibrium in the vertical direction,
supported by the gas pressure. As the model of the Be disk in
\gx, we assume for simplicity that the disk is isothermal at
$T_\mathrm{d}=0.6 T_\mathrm{eff}$ \citep{carciofi2006a}. We also
assume that the disk does not change over the X-ray activity period
and its density profile, $\rho(r,h)$, is cylindrically symmetric, i.e., 
depending on the distance, $r$, to the symmetry axis of the disk and
the height, $h$, above the disk plane:
\begin{equation}\label{eq:diskrho}
\rho(r,h) = \rho_0 \left( \frac{r}{R_*} \right)^{-n} \exp \left[ -\frac{h^2}{2H(r)^2} \right],
\end{equation}
where $\rho_0$ is the base density, i.e., the density at
$(r,h)=(R_*,0)$, $n$ is a constant that characterizes the radial
density distribution \citep[see, e.g.,][]{okazaki2013a}, $R_*$
is the radius of the Be star, and $H(r)$ is the vertical
scale-height given by
\begin{equation}
H(r) = \frac{c_\mathrm{s}}{\Omega_\mathrm{K}(r)} = \left(
\frac{k T_\mathrm{d} R_*^3}{\mu m_\mathrm{H} G M_*} \right)^{1/2}
\left( \frac{r}{R_*} \right)^{3/2},
\end{equation}
where $c_\mathrm{s}=(kT_\mathrm{d}/\mu m_\mathrm{H})^{1/2}$ is the
isothermal sound speed, with $\mu$ and $m_\mathrm{H}$ being the mean
molecular weight and the mass of the hydrogen atom, respectively,
and $\Omega_\mathrm{K}=(GM_*/R_*^3)^{1/2}$ is the Keplerian rotation
velocity with the mass, $M_*$, of the Be star. We adopt $R_* =
6\,\rsol$, $M_* = 10\,\msol$, and $T_\mathrm{eff} = 22000$\,K
consistent with a B2Vne star \citep{parkes1980a} and $\mu=0.62$ for
fully ionized plasma with cosmic abundances. We fixed the density
profile index $n=2.5$ as expected for a truncated disk as described
in the following.

In a binary like \gx, the Be disk is thought to be truncated at a
radius smaller than the binary separation at periastron. If the Be
disk is coplanar with the binary orbital plane, the truncation occurs
at a resonance radius, $r_{_\mathrm{T}}$, which mainly depends on the
orbital eccentricity and the disk viscosity
\citep{negueruela2001a,okazaki2001a}. For instance, for the viscosity
parameter $\alpha \sim 0.1$ and the above stellar and disk parameters
for \gx, the disk is truncated at the 6:1 resonance radius
($r_{_\mathrm{T}}/a=0.29$). If the Be disk is highly misaligned
as we consider for \gx, however, it has a significantly larger radius
than the coplanar disks, because of the weaker resonant torques, and
is likely to fill the Roche lobe of the Be star
\citep{lubow2015a,miranda2015b}. Given that there are large
uncertainties in the viscosity parameter (according to
\citealt{clark2001a} and \citealt{wisniewski2010a}, $\alpha \sim 0.1$,
while \citealt{carciofi2012a} assume $\alpha \sim 1$) and the
misalignment angle, we examine the disk obscuration effect for two
extreme disk cases, one truncated at the 6:1 resonance radius
($r_{_\mathrm{T}}/a=0.29$) and the other that fills the Roche lobe
radius averaged over the binary orbit ($r_{_\mathrm{T}}/a = 0.49$).
In the calculation of the latter radius, we applied the approximated
formula by \citet{eggleton1983a} to the averaged binary separation,
$d = a (1-e^2)^{1/2}$, as
\begin{equation}
r_{_\mathrm{T}} = \frac{0.49\,q^{2/3} d}{0.6\,q^{2/3} + \ln (1+q^{1/3})},
\label{eq:roche}
\end{equation}
where $q = M_*/M_\mathrm{X}$ is the binary mass ratio assuming the
canonical neutron star mass of $M_\mathrm{X} = 1.4\,\msol$. For radii
larger than the truncation radius, $r_{_\mathrm{T}}$, we assume that
the density distribution of the Be disk is
\begin{equation}\label{eq:diskrhotrunc}
  \rho(r > r_{_\mathrm{T}}, h) = \rho_{_\mathrm{T}}(h)
  \left(\frac{r}{r_{_\mathrm{T}}}\right)^{-m}
\end{equation}
with the density at the truncation radius $\rho_{_T}(h) =
\rho(r_{_\mathrm{T}}, h)$ as calculated after Eq.~\ref{eq:diskrho}.
The constant $m > n$ leads to a faster decrease of the density
profile than for radii $r <  r_{_\mathrm{T}}$ and we assume
$m = 10$, which is consistent with numerical results obtained by
\citet{okazaki2002a}.

Figure~\ref{fig:disksetup} shows the radial (a) and height (b)
dependence of the density profile as defined above in the case of
$r_{_\mathrm{T}}/a = 0.49$ with $a = 601$\,lt-s and
$\rho_0 =10^{-11}\,\mathrm{g}\,\mathrm{cm}^{-3}$.

\subsection{Density along the X-ray line of sight}
\label{sec:densitymodel}

\begin{figure*}
  \includegraphics[width=1.5\columnwidth]{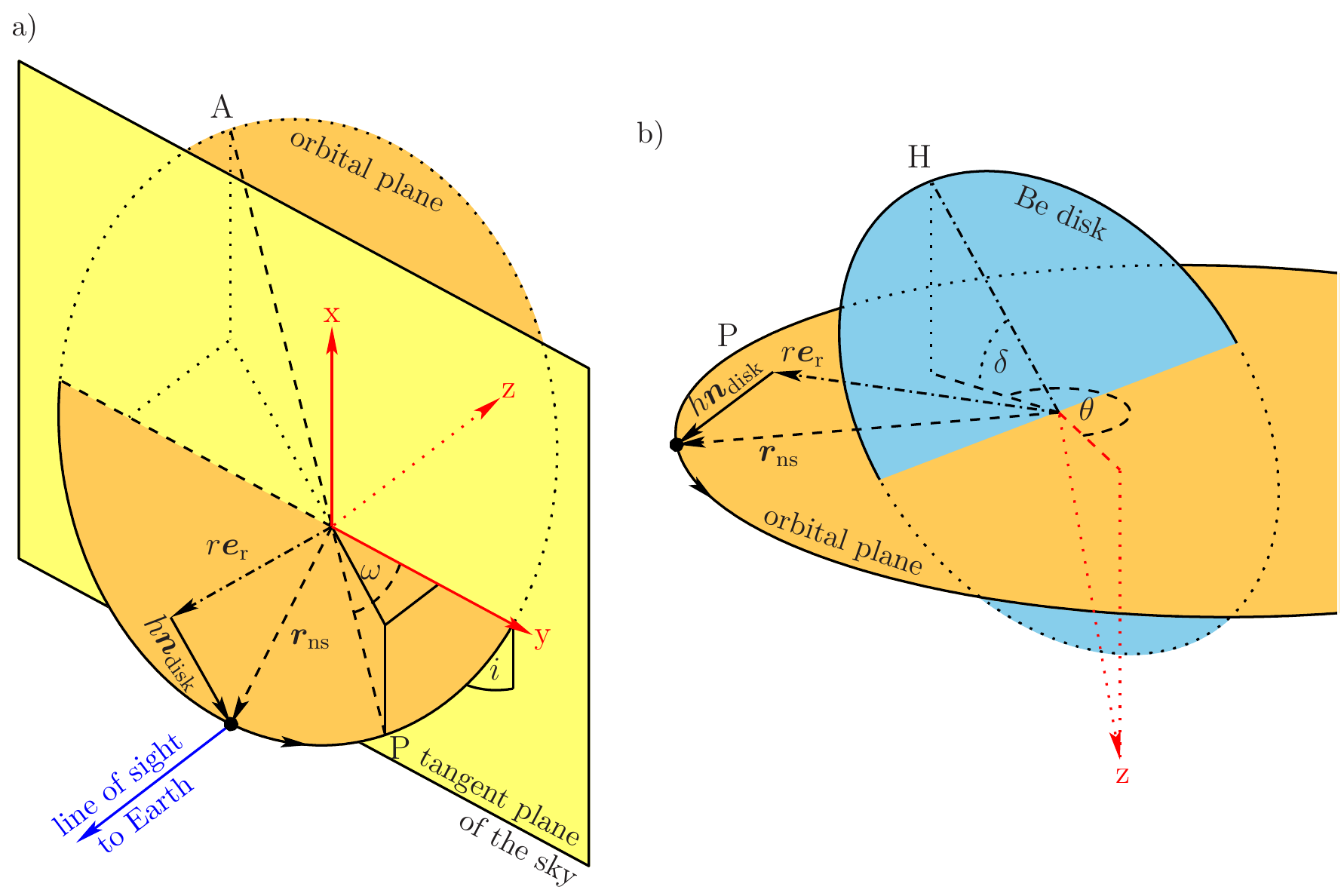}
  \caption{Definition of the reference frames as described in the text.
  \textbf{a)} The tangent plane of the sky (yellow) defines the cartesian
  reference frame ($x$, $y$, $z$; red arrows) with the $z$-axis and
  the line of sight (blue arrow) perpendicular to this plane. Dashed
  lines are within the orbital plane (orange). P and A denote the
  position of the periastron and apastron on the neutron star orbit,
  respectively. \textbf{b)} The Be disk (blue) is inclined with
  respect to the orbital plane (orange). The position vector of the
  neutron star ($\mathbfit{r}_\mathrm{ns}$) is decomposed using the
  unit vectors of the Be-disk plane ($\mathbfit{e}_\mathrm{r}$ and
  $\mathbfit{n}_\mathrm{disk}$). Dash-dotted lines are within the Be-disk
  plane. H marks the highest point of the disk above the orbital plane.
  Note that for clarity we show different orbital phases in plots a) and
  b).}
  \label{fig:refframe}
\end{figure*}

In order to calculate the absorption column density, $N_\mathrm{H}$, we
need to transform the line of sight to the neutron star, given in the
binary's reference frame, into the Be-disk's reference frame, i.e., into
cylindrical coordinates. Figure~\ref{fig:refframe} shows a sketch of the
following definitions. We define the origin of the binary's reference
frame to be fixed at the position of the Be-type companion star. The
xy-plane is equivalent to the tangent plane of the sky, i.e., the $z$-axis
is parallel to the line of sight to the companion and pointing away from
Earth. The position of the neutron star on its orbit is found by solving
Kepler's equation,
\begin{equation}\label{eq:kepler}
  E - e \sin E = M,
\end{equation}
for the eccentric anomaly, $E$. Here, $e$ is the eccentricity of
the orbit and
\begin{equation}\label{eq:meananomaly}
  M = 2 \pi \left(\frac{t - \tau}{P_\mathrm{orb}}\right)
\end{equation}
is the mean anomaly with the time of the observation, $t$, the time
of periastron passage, $\tau$, and the orbital period,
$P_\mathrm{orb}$ (see Table~\ref{tab:orbit} for the orbital
parameters of \gx). The position vector to the neutron star,
$\mathbfit{r}_\mathrm{ns} =
(x_\mathrm{ns}, y_\mathrm{ns}, z_\mathrm{ns})^\mathrm{T}$, is then
given by
\begin{equation}\label{eq:nspos}
  \begin{pmatrix}
    x_\mathrm{ns}\\y_\mathrm{ns}\\z_\mathrm{ns}
  \end{pmatrix} = \mathbfss{R}(\mathbfit{e}_\mathrm{y}, 90^\circ - i) \begin{pmatrix}
    0 \\
    a \cos \omega (\cos E - e) - b\sin E \sin \omega \\
    a \sin \omega (\cos E - e) + b\sin E \cos \omega
  \end{pmatrix}
\end{equation}
with the rotation matrix, $\mathbfss{R}$, around the unit vector along the
$y$-axis, $\mathbfit{e}_\mathrm{y}$, and the angle $90^\circ - i$ with the
inclination, $i$, of the orbital plane with respect to the tangent
plane of the sky. The semi-major axis, $a$, is found using the
measured value of $a \sin i$, $b = \sqrt{a^2 (1 - e^2)}$ is the
semi-minor axis, and $\omega$ is the argument of periastron.

In order to transform any vector in the binary's reference frame
into the reference frame of the Be disk with cylindrical
coordinates ($r$,$h$), we first calculate the normal vector,
$\mathbfit{n}_\mathrm{disk}$, of the disk by taking the inclination of
the orbit, $i$, the misalignment angle of the disk, $\delta$, and the
position angle of the disk, $\theta$, into account,
\begin{equation}\label{eq:disknormal}
  \mathbfit{n}_\mathrm{disk} = \mathbfss{R}(\mathbfit{n}_\mathrm{orb}, \theta)\,
  \mathbfss{R}(\mathbfit{e}_\mathrm{y}, 90^\circ - i + \delta)\,
  \mathbfit{e}_\mathrm{x}.
\end{equation}
Here, $\mathbfit{e}_\mathrm{x}$, is the unit vector along the $x$-axis (of
the binary's reference frame), and $\mathbfit{n}_\mathrm{orb}$ is the
normal vector of the orbital plane,
\begin{equation}\label{eq:orbitnormal}
  \mathbfit{n}_\mathrm{orb} = \mathbfss{R}(\mathbfit{e}_\mathrm{y},
  90^\circ - i)\,
  \mathbfit{e}_\mathrm{x}.
\end{equation}
Due to the fact that no enhancement event was detected in the earlier
outbursts of 2010, the Be disk was not in the X-ray line of sight, while
a strong occultation event occurred in 2011~May. A likely explanation is
that the Be disk in \gx is precessing. Since we expect the Be disk to
precess in the retrograde direction compared to the orbital movement of
the neutron star \citep{papaloizou1994a,papaloizou1995a}, we calculate
the position angle of the disk after
\begin{equation}\label{eq:diskposangle}
  \theta = \omega_\mathrm{disk} - \dot{\omega}_\mathrm{disk} (t -
  t_0),
\end{equation}
with the initial position angle $\omega_\mathrm{disk}$ at the time $t_0$
and the precession frequency, $\dot{\omega}_\mathrm{disk}$. Note that in
the context of the binary's reference frame defined above, the position
angle, $\theta$, is measured between the line of sight,
$\mathbfit{e}_\mathrm{z}$, and the highest point of the disk, both
projected onto the orbital plane (see Fig.~\ref{fig:refframe}). Finally,
the transformation, $T$, of any vector, $\mathbfit{r}$, given in the
binary's reference frame into the cylindrical coordinates of the
reference frame of the Be disk is
\begin{equation}\label{eq:coordtrafo}
  T{:}\quad\mathbfit{r} \rightarrow (r,h) = \left(
  \left|\mathbfit{r} - \left(\mathbfit{n}_\mathrm{disk} \cdot
  \mathbfit{r}\right)\,\mathbfit{n}_\mathrm{disk}\right|,~
  \mathbfit{n}_\mathrm{disk} \cdot \mathbfit{r}\right).
\end{equation}

\begin{table}
  \caption{Parameters of the Be-disk occultation model as defined in
    Eqs.~\ref{eq:kepler}--\ref{eq:model}. See the text for a
    detailed description and for the assumptions of fixing certain
    parameters.}
  \centering
  \begin{tabular}{lll}
  \hline
  Be-star radius & $R_* = 6\,\rsol$ & (fixed) \\
  Be-star mass   & $M_* = 10\,\msol$ & (fixed) \\
  Be-disk temperature & $T_\mathrm{eff} = 22000$\,K & (fixed) \\
  Be-disk truncation radius (see text) & $r_{_\mathrm{T}}$ & (fixed) \\
  Be-disk base density & $\rho_0$ \\
  Density profile index & $n = 2.5$ & (fixed) \\
  Density profile index (truncated) & $m = 10$ & (fixed) \\
  Orbital parameters (see Table~\ref{tab:orbit}) & $P_\mathrm{orb}$,
  $a \sin i$, $e$, $\tau$, $\omega$ & (fixed) \\
  Orbit inclination & $i$ \\
  Be-disk inclination & $\delta$ \\
  Be-disk position angle at $t_0$ & $\omega_\mathrm{disk}$ \\
  Be-disk precession frequency & $\dot{\omega}_\mathrm{disk}$ \\
  Precession reference time & $t_0 = \mathrm{MJD}~55690$ & (fixed) \\
  Foreground absorption & $N_\mathrm{H,frgrd}$ \\
  \hline
  \end{tabular}
  \label{tab:params}
\end{table}

The absorption column density is finally found by computing the
integral of the Be-disk density, $\rho$
(Eqs.~\ref{eq:diskrho}--\ref{eq:diskrhotrunc}), along the line of
sight, i.e., along the $z$-axis up to the neutron star's position:
\begin{equation}\label{eq:model}
  N_\mathrm{H} = N_\mathrm{H,frgrd} + \frac{1}{m_\mathrm{H}}
  \int_{-\infty}^{z_\mathrm{ns}} \rho(T((x_\mathrm{ns}, y_\mathrm{ns},
  z^\prime)^\mathrm{T})) \mathrm{d}z^\prime 
\end{equation}
Here, $N_\mathrm{H,frgrd}$ accounts for constant interstellar
foreground absorption. Table~\ref{tab:params} summarizes the fixed
and free parameters of the full model.

\subsection{Modelling the observed absorption column densities}

Initial attempts to apply the model defined in Eq.~\ref{eq:model} to
the observed $N_\mathrm{H}$ evolution over time shown in
Fig.~\ref{fig:parsphi}b revealed several issues. First, finding the
best-fit using a commonly used $\chi^2$-minimization is complicated by
several local minima within the $\chi^2$-landscape, where the actual
observed enhancement event in 2011~May is not modelled at all.
Furthermore, $\chi^2_\mathrm{red}$ is much larger than unity even
during outbursts where no occultation event is detected due to a
significant scattering of the observed $N_\mathrm{H}$ values (see
discussion below). Finally, from the calculation of the disk's normal
vector, $\mathbfit{n}_\mathrm{disk}$ (Eq.~\ref{eq:disknormal}), we expect a
parameter degeneracy between the orbit inclination, $i$, and the
Be-disk inclination, $\delta$.

In order to solve these issues, we applied a Bayesian analysis of the
model to the $N_\mathrm{H}$ evolution in form of a Markov chain Monte
Carlo (MCMC) sampling approach after \citet{goodman2010a}. We used the
\texttt{emcee} algorithm as implemented by \citet{foreman-mackey2013a}
and ported into ISIS by M.A.~Nowak, which is distributed via the
ISISscripts\footnote{\url{http://www.sternwarte.uni-erlangen.de/isis}}.
The advantages of this approach are that parameter degeneracies are
automatically error propagated and the algorithm always provides the
most probable answer even in cases of a bad goodness of the fit. The
result of an \texttt{emcee} run is the probability distribution for
each free parameter of the model used. The most probable parameters,
i.e., the best-fit parameters, correspond to the maxima found in the
probability distributions. These are sampled by so-called
\textit{walkers}, which move within the parameter space and are
distributed uniformly at the beginning.

For each possible truncation radius ($r_{_\mathrm{T}}/a = 0.49$ or 0.29,
see Sect.~\ref{sec:diskprofile}), we have performed an \texttt{emcee}
run with 1\,000 walkers per free parameter and 10\,000 iteration
steps. The resulting parameter chains were investigated on convergence,
i.e, the acceptance rate is stable around 0.25 after ${\sim}500$
iteration steps, which is in the range of 0.2--0.5 as expected in
convergence \citep{foreman-mackey2013a}. Thus, we ignored the first 500
iteration steps in the following analysis of the parameter
distributions.

\begin{figure}
  \includegraphics[width=\columnwidth]{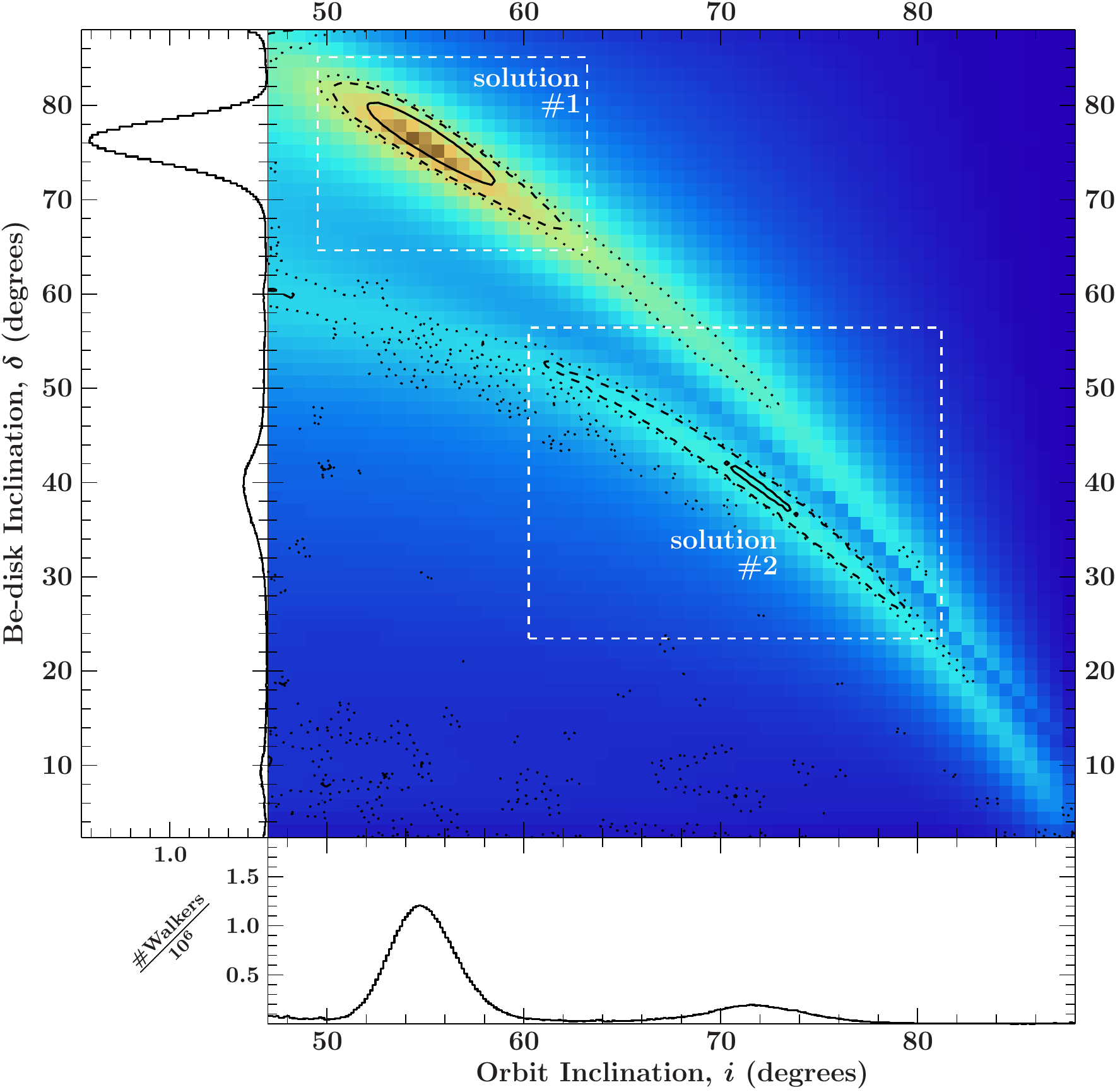}
  \caption{2D-probability contours (black) between the orbit
  inclination, $i$, and the Be-disk inclination, $\delta$, with
  respect to the orbital plane for the case $r_{_\mathrm{T}}/a = 0.49$.
  The line style indicates the 68\% (solid), 90\% (dashed), and 99\%
  (dotted) confidence level. In order to check the \texttt{emcee}
  result, the underlying colour map shows the corresponding $\chi^2$
  at each point, calculated independently of the \texttt{emcee} run.
  The vertical and horizontal histograms correspond to the
  1D-probability distributions of both parameters, i.e., summing up
  the columns or rows of the map, respectively. The white, dashed
  boxes mark the regions which have been used to split the parameter
  chains into the two distinct solutions (\#1 and \#2).}
  \label{fig:incconfmap}
\end{figure}

Almost all probability distributions, which are found by sorting the
corresponding parameter chain into a histogram, show two or three
distinct maxima, i.e., multiple possible solutions. These solutions
are best seen in the 2D-probability distribution (i.e., similar to a
$\chi^2$ contour map) between the orbit inclination, $i$, and the
Be-disk inclination, $\delta$. Figure~\ref{fig:incconfmap} shows this
distribution for a Be-disk truncation radius of $r_{_\mathrm{T}}/a =
0.49$. The peak with the highest probability (solution \#1)
corresponds to a high disk misalignment angle, $\delta$, and a
moderate orbit inclination, $i$, while its the other way around for
the second highest peak (solution \#2). Furthermore, the degeneracy
between these parameters as expected from Eq.~\ref{eq:disknormal} is
visible as diagonal valleys in the figure. In the case of
$r_{_\mathrm{T}}/a = 0.29$ even a third solution appears (solution \#3).

\begin{table*}
  \caption{Most probable parameters for both extreme cases of
  Be-disk truncation and for all solutions discovered in the
  parameter space. See Table~\ref{tab:params} for a brief
  description of the model parameters.}
  \renewcommand{\arraystretch}{1.15}
\begin{tabular}{lllllll}
\hline
\multicolumn{2}{l}{$r_{_\mathrm{T}}/a$ [solution no.]} & $0.49$ [\#1] & $0.49$ [\#2] & $0.29$ [\#1] & $0.29$ [\#2] & $0.29$ [\#3]\\[1mm]
Parameter & (unit) \\\hline
$i$ & (degree) & $54.7^{+1.9}_{-1.5}$ & $71.7^{+2.6}_{-2.3}$ & $67.1^{+0.8}_{-0.8}$ & $82.5^{+1.1}_{-1.4}$ & $54.5^{+1.7}_{-1.7}$ \\
$\delta$ & (degree) & $76.3^{+2.0}_{-2.5}$ & $40^{+4}_{-5}$ & $62.7^{+1.3}_{-1.5}$ & $17.8^{+3.1}_{-2.8}$ & $81.9^{+2.0}_{-2.3}$ \\
$\omega_\mathrm{disk}$ & (degree) & $121.5^{+0.8}_{-0.8}$ & $89.4^{+2.0}_{-1.5}$ & $138.0^{+1.9}_{-1.8}$ & $74.5^{+2.4}_{-2.6}$ & $131.9^{+2.5}_{-2.6}$ \\
$\dot{\omega}_\mathrm{disk}$ & (degree\,yr$^{-1}$) & $190^{+4}_{-5}$ & $250^{+5}_{-6}$ & $135^{+34}_{-13}$ & $271^{+8}_{-10}$ & $176^{+5}_{-5}$ \\
$\log(\rho_0 /$g\,cm$^{-3})$ &  & $-10.29^{+0.04}_{-0.05}$ & $-10.62^{+0.06}_{-0.07}$ & $-10.22^{+0.06}_{-0.06}$ & $-10.81^{+0.06}_{-0.05}$ & $-9.37^{+0.16}_{-0.13}$ \\
$N_\mathrm{H,frgrd}$ & ($10^{22}\,$cm$^{-2}$) & $3.26^{+0.13}_{-0.13}$ & $3.22^{+0.18}_{-0.15}$ & $3.79^{+0.11}_{-0.11}$ & $3.64^{+0.29}_{-0.13}$ & $3.75^{+0.14}_{-0.12}$ \\
$\chi^2$ / d.o.f. &  & 252.03 / 63 & 303.11 / 63 & 313.08 / 63 & 290.51 / 63 & 307.33 / 63 \\
\hline
\end{tabular}
  \label{tab:bestfit}
\end{table*}

\begin{figure*}
  \includegraphics[width=\textwidth]{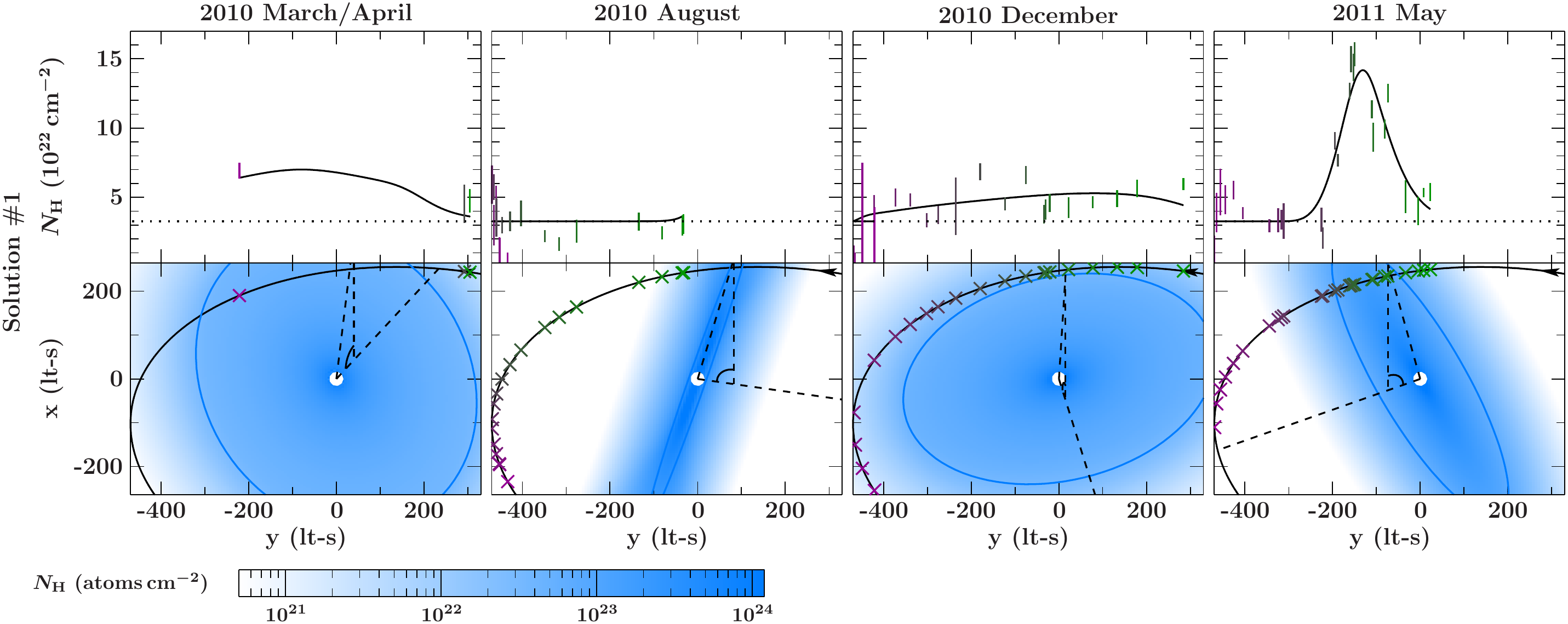}\\[1mm]
    \includegraphics[width=\textwidth]{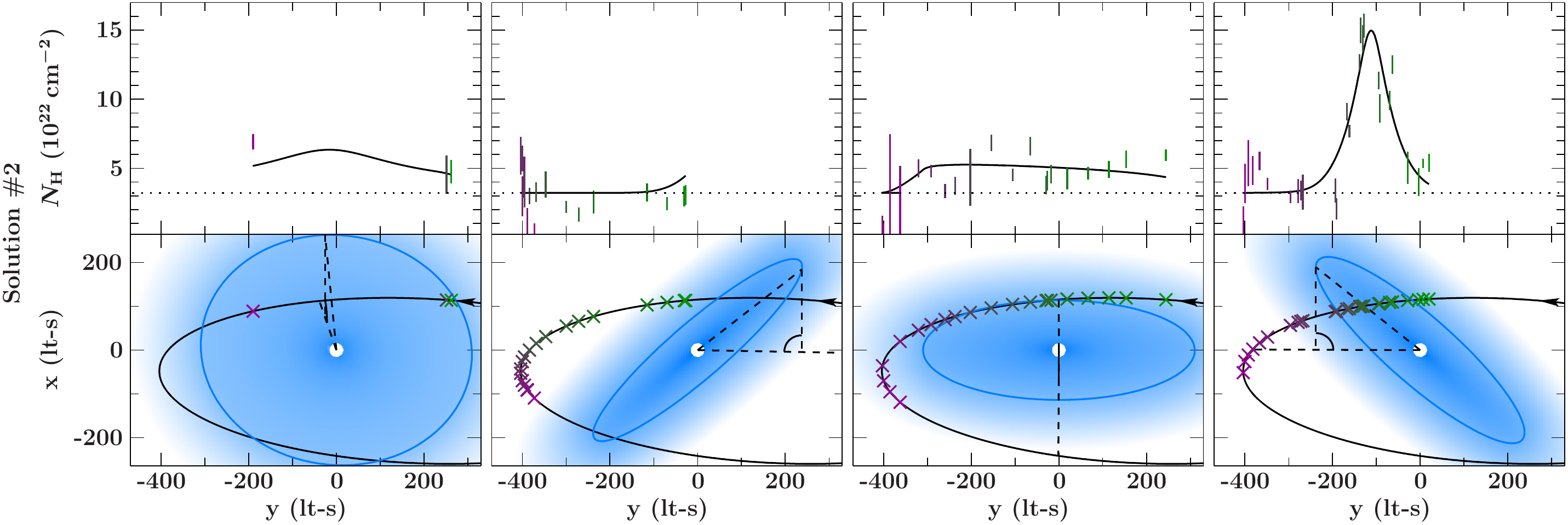}
    \caption{Absorption column and projected geometry of the Be disk
      for the case of $r_{_\mathrm{T}}/a = 0.49$. The upper row shows
      the results for solution \#1, the lower row presents the results
      for solution \#2. Results for $r_{_\mathrm{T}}/a = 0.29$ are not
      shown as they are very similar. The top panels in each subfigure
      show measured (green/purple) and modelled (black) absorption
      column densities, $N_\mathrm{H}$, for all four outbursts of \gx
      (each column represents one outburst). Colours of data points
      indicate the time of the observation during a certain outburst
      (from green to purple). The dotted line is the most probable
      interstellar foreground absorption, $N_\mathrm{H,frgrnd}$ (see
      Table~\ref{tab:bestfit}). The bottom panels in each subfigure
      illustrate the modelled Be-disk geometry and neutron star
      position as seen from Earth, i.e., the components of the binary
      are projected onto the tangent plane of the sky. The white
      circle in the centre of the reference frame is the Be star, the
      crosses mark the positions of the neutron star along its orbit
      (black line) during all observations. The blueish region is
      proportional to the $N_\mathrm{H}$ along the line sight through
      the Be disk, and the blue circle marks the truncation radius,
      $r_{_\mathrm{T}}/a = 0.49$, of the Be disk within the disk
      plane. The highest point on the disk's rim above the orbital
      plane is connected with the central Be star by one dashed line
      and to the orbital plane by another line perpendicular to the
      latter. The dashed line from the Be star to the neutron star's
      orbit marks the orbital phase of the highest point of the disk,
      i.e., the position angle of the disk,
      $\omega_\mathrm{disk}$.} \label{fig:bestfit}
\end{figure*}

In order to reduce the influence of these solutions on the
1D-probability distributions, which are used to derive the parameter
values and uncertainties, we have split the parameter chains into areas
around each solution. Fig.~\ref{fig:incconfmap} shows these areas for
the case of $r_{_\mathrm{T}}/a = 0.49$. In this way, the number of peaks
in the probability distribution of each parameter reduces to one, i.e.,
the solution is unique and we are able to provide the final parameter
values for all possible solutions and the two extreme cases for the
truncation radius. The most probable parameter value is found by
determining the peak's position using a polynomial fit around the
maximum. The lower and upper confidence levels are determined such that
68\% of the peak's total area is within this confidence interval.
Thereby, a linear continuum has been subtracted from the probability
distribution in order to investigate the area of the peak only.
Table~\ref{tab:bestfit} lists the resulting most probable parameters and
uncertainties for each investigated truncation radius and solution found
in the parameter space. For the case of $r_{_\mathrm{T}}/a = 0.49$, we
compare in Fig.~\ref{fig:bestfit} the modelled evolution of
$N_\mathrm{H}$ for both solutions with the data. We also show the
resulting geometries of the binary and Be disk projected onto the
tangent plane of the sky, i.e., as seen from Earth.

\section{Discussion \& Conclusions}
\label{sec:conclude}

In this paper we have presented a study of the $N_\mathrm{H}$
behaviour of \gx during four outbursts in terms of an occultation by a
precessing Be-star disk. The most probable parameters listed in
Table~\ref{tab:bestfit} correspond to the different solutions
discovered in the parameter space.

Formally, there are five solutions across the two truncation radii ---
two for $r_{_\mathrm{T}}/a=0.49$ and three for
$r_{_\mathrm{T}}/a=0.29$. However, these really reflect only three
basic physical scenarios, which we refer to as solutions \#1, \#2, and
\#3.  Solution \#1 and \#2 are found for both $r_{_\mathrm{T}}/a=0.49$
and $r_{_\mathrm{T}}/a=0.29$, while Solution \#3 only appears in the
$r_{_\mathrm{T}}/a=0.29$ case. Solution \#1 corresponds to a system
with moderate orbital inclination and a highly-misaligned,
high-density Be-disk which precesses relatively slowly. Solution \#2
has a higher inclination, but a less-misaligned, lower-density Be-disk
precessing more quickly. Finally, solution \#3 has a very highly
misaligned, very high-density Be-disk in a system with a precession
frequency and inclination comparable to solution \#1.

We have investigated additional truncation radii and found that the
parameters of solutions \#1 and \#2 change nonlinearly with truncation
radius. As the truncation radius is decreased, solution \#3 migrates
into the parameter space, starting at very high disk misalignment
angles (the upper left corner in Fig.~\ref{fig:incconfmap}). Which
solution is the most realistic cannot be decided easily, as discussed
in the following.

The best nominal solution in terms of $\chi^2$ is a large Be disk with
a high misalignment angle ($r_{_\mathrm{T}}/a=0.49$, \#1, first column
in Table~\ref{tab:bestfit}). From the statistical point of view this
solution is, however, still not acceptable ($\chi^2_\mathrm{red} =
4.27$). The unacceptable high $\chi^2$ values may have several
reasons: first, our simple model assumes a rigid and cylindrically
symmetric Be disk. In particular, we ignore any structures within the
Be disk, such as warping and spiral density waves, which are known to
be present due to the tidal interaction with the neutron star
(see, e.g., \citealt{okazaki2002a} and \citealt{martin2011a}).
Such a behaviour might explain the observed scattering of the
$N_\mathrm{H}$ values, especially during the 2010~December outburst.
Here, our most probable model suggests that the line of sight crosses
the outermost parts of the Be disk near the truncation radius
(Fig.~\ref{fig:bestfit}), where the tidal effects are expected to be
most prominent as discussed for the BeXRB 4U~0115$+$634 by
\citet{negueruela2001a} and \citet{reig2007a} and for A0535+262 by
\citet{moritani2013a}. Secondly, the $N_\mathrm{H}$ measurements
itself might be influenced by systematic effects, such as the Xe
L-edge, which \citetalias{rothschild2017a} had to introduce in order
to model calibration uncertainties in \rxte-PCA. Further evidence for
these systematics is that the \texttt{emcee} runs prefer a large
interstellar foreground absorption of $N_\mathrm{H,frgrd} \sim 3.6
\times 10^{22}\,\mathrm{cm}^{-2}$. Although \gx is located in
direction of the Coalsack nebula, \textsl{Suzaku} observations during
its 2010~August and 2012~January outbursts revealed $N_\mathrm{H} \sim
1 \times 10^{22}\,\mathrm{cm}^{-2}$ \citep{jaisawal2016a}, which is
consistent with the foreground absorption found in 21\,cm surveys
\citep{kalberla2005a,dickey1990a}. Assuming a systematic uncertainty
of $2 \times 10^{22}\,\mathrm{cm}^{-2}$ to be consistent to these
alternative foreground absorption measurements indeed results in a
$\chi^2_\mathrm{red}$ around unity. This does not, however, affect the
significance of the occultation event in 2011~May.

We have found that the base density of the Be disk, $\rho_0$, which is
mainly determined by the measured $N_\mathrm{H}$ values during the
occultation event in 2011~May, is on the order of
$10^{-11}$--$10^{-10}\,\mathrm{g}\,\mathrm{cm}^{-3}$. This result is in
very good agreement with the commonly accepted value of
$10^{-11}\,\mathrm{g}\,\mathrm{cm}^{-3}$ \citep[e.g.,][]{okazaki2013a}
for disks of isolated Be stars. Although the Be disks in binaries are
expected to be approximately twice as dense as compared to isolated
stars \citep{zamanov2001a}, the solutions with densities significantly
above $10^{-10}\,\mathrm{g}\,\mathrm{cm}^{-3}$ would be unlikely, such as
solution \#3 with $r_{_\mathrm{T}}/a=0.29$.

The precession period, $\dot{\omega}_\mathrm{disk}$, of solution \#2
is about 30\%--100\% faster than that of solution \#1. This difference
decreases with increasing truncation radius. According to
\citet{larwood1998a}, the precession period, $P_\mathrm{disk}$, due to
tidal forces on a misaligned disk is
\begin{equation}\label{eq:Pdisk}
  P_\mathrm{disk} = \frac{7}{3} \left(\frac{1+q}{q^2}\right)^{1/2}
  \left(\frac{a}{r_{_\mathrm{T}}}\right)^{3/2}
  \frac{P_\mathrm{orb}}{\cos \delta}
\end{equation}
with $q = M_\mathrm{X}/M_*$. Using the resulting disk misalignment
angle, $\delta$, of solution \#1 for the truncation radius
$r_{_\mathrm{T}}/a = 0.49$ we find \mbox{$P_\mathrm{disk} \sim
  80$\,yr}, which is ${\times}42$ longer than the \texttt{emcee}
result of $360^\circ / \dot{\omega}_\mathrm{disk} \sim 1.9$\,yr. Using
other \texttt{emcee} solutions does not weaken this disagreement
significantly. Since the material forming the Be disk originates from
the central Be star, the disk's precession might be induced by an
intrinsically precessing Be star. This precession is the result of
tidal forces on an oblate star such as a rapidly rotating Be
star. Using the theoretical investigations by \citet{kopal1959a} and
\citet{alexander1976a}, we have estimated the precession period of the
Be star in \gx to be on the order of 5000\,yr (see
appendix~\ref{app:precess} for the detailed calculation). This rules
out Be star precession as the origin of the fast Be disk precession we
have derived. However, in case of eccentric orbits, as usually present
in BeXRBs, the eccentric Kozai-Lidov (KL) mechanism has to be taken
into account for the torque on misaligned disks as calculated after
Eq.~\ref{eq:Pdisk}. Following the formulation by \citet{naoz2013a} and
\citet{naoz2016a}, we have numerically investigated the evolution of
Be-disk particle orbits with initial radii between $r_{_\mathrm{T}}/a
= 0.49$ and $r_{_\mathrm{T}}/a = 0.29$ and similar misalignment
angles, $\delta$, as listed in Table~\ref{tab:bestfit}. As a result
from the KL mechanism, the Be-disk particles precess much faster
around the binary's orbital momentum vector than compared to the
circular approximation of Eq.~\ref{eq:Pdisk}. For a Be-disk radius of
$r_{_\mathrm{T}}/a = 0.29$ the precessing time-scale is ${\sim}10$\,yr
and decreases to 4--5\,yr for $r_{_\mathrm{T}}/a = 0.49$ for all
misalignment angles, which is in agreement with our best-fit within a
factor of 2--3. A very similar time-scale was found in A~0535+262 by
\citet{moritani2013a}, who performed an optical spectroscopic
monitoring of its Be-type companion star. From the measured radial
velocity curve of an enhanced component in the H$\alpha$ line profile,
they have derived a period of ${\sim}674$\,d, which they interpret as
the precession period of a warped Be-disk.

\begin{figure}
  \includegraphics[width=\columnwidth]{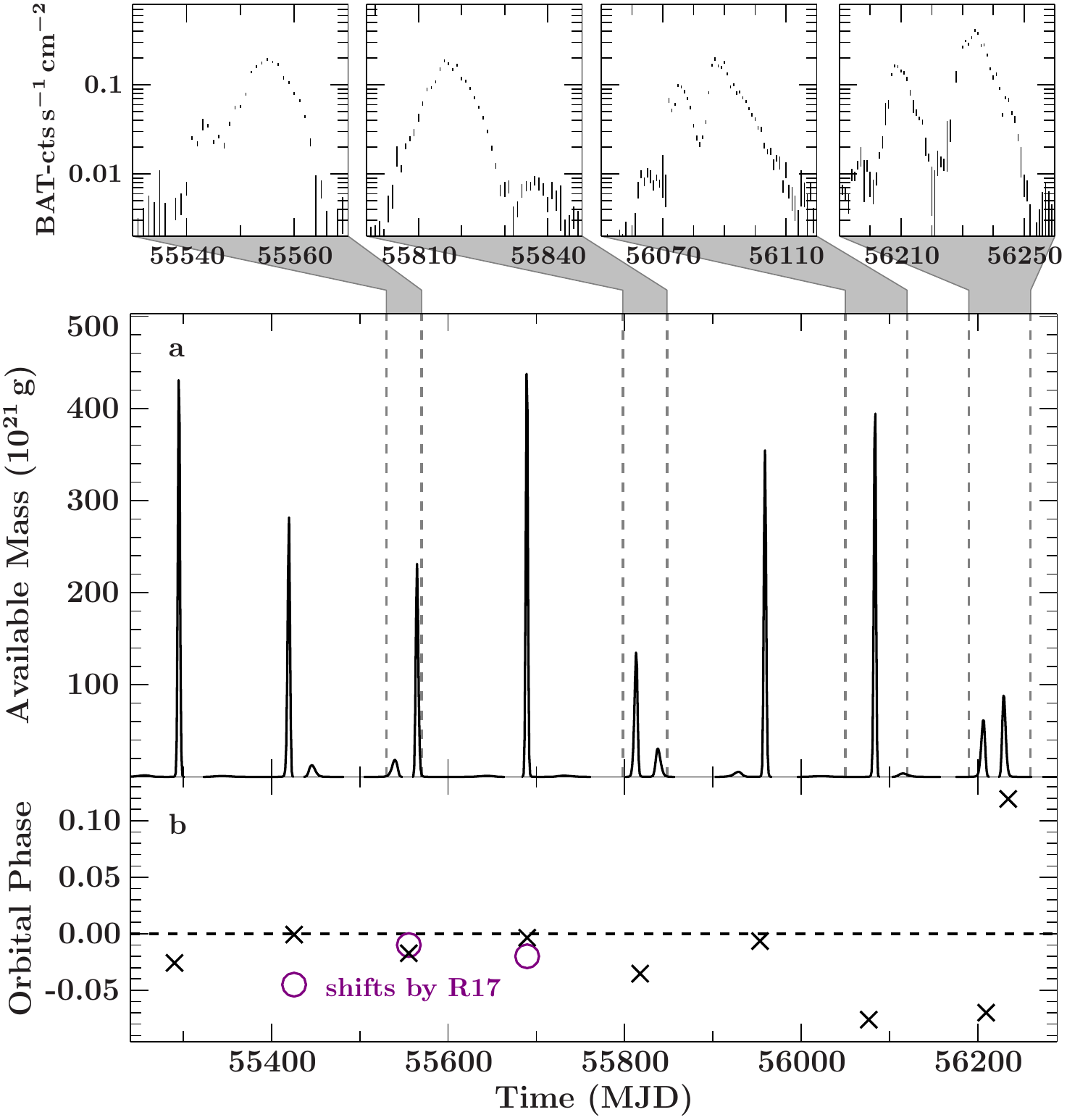}
  \caption{\textbf{a)} The available Be-disk mass within the Roche
    radius of the neutron star as a function of time for
    $r_{_\mathrm{T}}/a = 0.49$ and solution \#1. The plots on top show
    the observed \textsl{Swift}-BAT light curve of \gx around some
    peaks in the predicted available mass. \textbf{b)} The orbital
    phases (black crosses) for the maxima in a) around the observed
    outbursts. \citetalias{rothschild2017a} introduced shifts in
    orbital phase (purple circles) in order to match the falling edges
    of the outburst light curves.}
  \label{fig:accretionrate}
\end{figure}

The misalignment angles of the Be disk found by our analysis are in
the range of $15^\circ < \delta < 85^\circ$. The most probable
solution ($r_{_\mathrm{T}}/a = 0.49$, \#1) points, however, to a large
misalignment angle of around $80^\circ$. As discussed by
\citet{brandt1995a} a high-velocity supernova kick of the neutron
star's progenitor might result in a large misalignment angle. In fact,
they suggested that the Be disk in BeXRBs are typically not aligned
with the orbital plane. An inclined Be disk in \gx with respect to the
orbital plane was also proposed by \citet{postnov2015a} as one
possible explanation for the occurrence of the two observed
double-peaked outbursts of the source in 2010~June and 2010~October.
We have investigated this idea further and calculated the available
Be-disk mass within the Roche radius of the neutron star as calculated
after Eq.~\ref{eq:roche} for the mass ratio $q = M_\mathrm{X}/M_*$ and
$d = a (1-e^2)/(1+e \cos(f))$ with the true anomaly, $f$, of the
neutron star's position. As shown in Fig.~\ref{fig:accretionrate}a the
resulting mass within the Roche radius around the observed
double-peaked outburst of \gx at MJD~56220 indeed shows two distinct
maxima. During this outburst, the orbital phase of the highest point
of the disk is near zero, i.e., at the periastron passage of the
neutron star. Thus, the neutron star orbit crosses the disk plane
twice. This strengthens the idea that double-peaked outbursts are due
to a misaligned Be disk, which moves into the orbit of the neutron
star as proposed by \citet{postnov2015a} for \gx and by
\citet{okazaki2013a} in general. Furthermore, the model predicts weak
leading and trailing flares to the periastron outbursts, which are
indeed observed at MJD~55540 and MJD~55830 in the \textsl{Swift}-BAT
light curve (shown on top of Fig.~\ref{fig:accretionrate}a). Our model
fails to explain the observed double-peaked outburst around MJD~56080
and the available mass during each outburst does not scale with the
observed count rate of \gx as measured with \textsl{Swift}-BAT. The
available mass can be modified, however, due to a warped disk, density
fluctuations, or tidal streams. The latter results from overflowing
gas of the Be disk close to periastron, which gets unbound as soon as
the Roche radius of the companion star decreases below the disk's
truncation radius, which is the case for $r_{_\mathrm{T}}/a \gtrsim
0.30$ in \gx. Finally, deriving the mass accretion rate and, thus, the
luminosity from the available mass requires further geometrical and
hydrodynamical investigations, which is beyond the scope of this
paper.

Figure~\ref{fig:accretionrate}b shows the orbital phases at which the
available mass within the neutron star's Roche lobe was at a maximum
(including both maxima for the double-peaked outburst around
MJD~56220). The orbital phases scatter around the periastron passage
between $-0.08$ and $+0.12$. For the 2010 August, 2010 December, and
2011 May outbursts of \gx, \citetalias{rothschild2017a} noticed
orbital phase shifts between the falling edges of these outbursts.
Interestingly, our derived orbital phases match those of
\citetalias{rothschild2017a} to some extent
(Fig.~\ref{fig:accretionrate}b, purple circles).

In summary, a precessing and inclined Be disk explains the observed
$N_\mathrm{H}$-evolution of \gx and supports the occurrence of
double-peaked outbursts and phase shifts of the outburst light curves.
Our model favours a large Be disk truncated close to the Roche radius
for the averaged binary separation. The fast disk precession found can
be explained by the eccentric Kozai-Lidov mechanism. However, some
issues still remain such as the unacceptable goodness of the fit, the
high number of possible solutions, and mismatches between some predicted
and observed double-peaked outbursts. We stress that until these issues
are solved any numbers derived from our analysis should be taken with
care and understood in a qualitative way. Increasing the data sample of
observed column densities towards \gx in combination with optical
observation of its companion, in particular H$\alpha$ line profiles,
would help solving these issues in the future.

\section*{Acknowledgements}

MK acknowledges support by the Bundesministerium f\"ur Wirtschaft und
Technologie under Deutsches Zentrum f\"ur Luft- und Raumfahrt grants
50OR1113 and 50OR1207 and by the European Space Agency under contract
number C4000115860/15/NL/IB. SMN acknowledges support by research
project ESP2016-76683-C3-1-R. JMT acknowledges research grant
ESP2014-53672-C3-3P. We thank the Deutscher Akademischer
Austauschdienst (DAAD) for funding JC through the Research Internships
in Science and Engineering (RISE). Parts of this work can be found in
the Ph.D.\ thesis of \citet{mueller2013a}. All figures within this
paper were produced using the \texttt{SLXfig} module, which was
developed by John E. Davis. We thank the \rxte-team for accepting and
performing our observations of \gx. Finally, we acknowledge the
comments by the referee, which helped improving the content of our
paper.

\bibliographystyle{mnras}


\begin{thebibliography}{}
\makeatletter
\relax
\def\mn@urlcharsother{\let\do\@makeother \do\$\do\&\do\#\do\^\do\_\do\%\do\~}
\def\mn@doi{\begingroup\mn@urlcharsother \@ifnextchar [ {\mn@doi@}
  {\mn@doi@[]}}
\def\mn@doi@[#1]#2{\def\@tempa{#1}\ifx\@tempa\@empty \href
  {http://dx.doi.org/#2} {doi:#2}\else \href {http://dx.doi.org/#2} {#1}\fi
  \endgroup}
\def\mn@eprint#1#2{\mn@eprint@#1:#2::\@nil}
\def\mn@eprint@arXiv#1{\href {http://arxiv.org/abs/#1} {{\tt arXiv:#1}}}
\def\mn@eprint@dblp#1{\href {http://dblp.uni-trier.de/rec/bibtex/#1.xml}
  {dblp:#1}}
\def\mn@eprint@#1:#2:#3:#4\@nil{\def\@tempa {#1}\def\@tempb {#2}\def\@tempc
  {#3}\ifx \@tempc \@empty \let \@tempc \@tempb \let \@tempb \@tempa \fi \ifx
  \@tempb \@empty \def\@tempb {arXiv}\fi \@ifundefined
  {mn@eprint@\@tempb}{\@tempb:\@tempc}{\expandafter \expandafter \csname
  mn@eprint@\@tempb\endcsname \expandafter{\@tempc}}}

\bibitem[\protect\citeauthoryear{{Alexander}}{{Alexander}}{1976}]{alexander1976a}
{Alexander} M.~E.,  1976, Ap\&SS, 45, 105

\bibitem[\protect\citeauthoryear{{Brandt} \& {Podsiadlowski}}{{Brandt} \&
  {Podsiadlowski}}{1995}]{brandt1995a}
{Brandt} N.,  {Podsiadlowski} P.,  1995, MNRAS, 274, 461

\bibitem[\protect\citeauthoryear{{Carciofi} \& {Bjorkman}}{{Carciofi} \&
  {Bjorkman}}{2006}]{carciofi2006a}
{Carciofi} A.~C.,  {Bjorkman} J.~E.,  2006, ApJ, 639, 1081

\bibitem[\protect\citeauthoryear{{Carciofi}, {Bjorkman}, {Otero}, {Okazaki},
  {{\v S}tefl}, {Rivinius}, {Baade}  \& {Haubois}}{{Carciofi}
  et~al.}{2012}]{carciofi2012a}
{Carciofi} A.~C.,  {Bjorkman} J.~E.,  {Otero} S.~A.,  {Okazaki} A.~T.,  {{\v
  S}tefl} S.,  {Rivinius} T.,  {Baade} D.,   {Haubois} X.,  2012, ApJ, 744, L15

\bibitem[\protect\citeauthoryear{{Clark}, {Tarasov}, {Okazaki}, {Roche}  \&
  {Lyuty}}{{Clark} et~al.}{2001}]{clark2001a}
{Clark} J.~S.,  {Tarasov} A.~E.,  {Okazaki} A.~T.,  {Roche} P.,   {Lyuty}
  V.~M.,  2001, A\&A, 380, 615

\bibitem[\protect\citeauthoryear{Dickey \& Lockman}{Dickey \&
  Lockman}{1990}]{dickey1990a}
Dickey J.~M.,  Lockman F.~J.,  1990, ARA\&A, 28, 215

\bibitem[\protect\citeauthoryear{{Eggleton}}{{Eggleton}}{1983}]{eggleton1983a}
{Eggleton} P.~P.,  1983, ApJ, 268, 368

\bibitem[\protect\citeauthoryear{{Finger} et~al.,}{{Finger}
  et~al.}{2009}]{finger2009a}
{Finger} M.~H.,  et~al., 2009, ArXiv Astrophysics e-prints

\bibitem[\protect\citeauthoryear{{Foreman-Mackey}, {Hogg}, {Lang}  \&
  {Goodman}}{{Foreman-Mackey} et~al.}{2013}]{foreman-mackey2013a}
{Foreman-Mackey} D.,  {Hogg} D.~W.,  {Lang} D.,   {Goodman} J.,  2013, PASP,
  125, 306

\bibitem[\protect\citeauthoryear{{Goodman} \& {Weare}}{{Goodman} \&
  {Weare}}{2010}]{goodman2010a}
{Goodman} J.,  {Weare} J.,  2010, Commun. Appl. Math. Comput. Sci., 5, 65

\bibitem[\protect\citeauthoryear{{Hanuschik}}{{Hanuschik}}{1996}]{hanuschik1996a}
{Hanuschik} R.~W.,  1996, A\&A, 308, 170

\bibitem[\protect\citeauthoryear{{Hemphill}, {Rothschild}, {Markowitz},
  {F{\"u}rst}, {Pottschmidt}  \& {Wilms}}{{Hemphill}
  et~al.}{2014}]{hemphill2014a}
{Hemphill} P.~B.,  {Rothschild} R.~E.,  {Markowitz} A.,  {F{\"u}rst} F.,
  {Pottschmidt} K.,   {Wilms} J.,  2014, ApJ, 792, 14

\bibitem[\protect\citeauthoryear{{Houck} \& {Denicola}}{{Houck} \&
  {Denicola}}{2000}]{houck2000a}
{Houck} J.~C.,  {Denicola} L.~A.,  2000, in Manset N.,  Veillet C.,   Crabtree
  D.,  eds,  Astron. Soc. of the Pacific Conf. Series Vol. 216, Astronomical
  Data Analysis Software and Systems IX. p.~591

\bibitem[\protect\citeauthoryear{{Jahoda}, {Markwardt}, {Radeva}, {Rots},
  {Stark}, {Swank}, {Strohmayer}  \& {Zhang}}{{Jahoda}
  et~al.}{2006}]{jahoda2006a}
{Jahoda} K.,  {Markwardt} C.~B.,  {Radeva} Y.,  {Rots} A.~H.,  {Stark} M.~J.,
  {Swank} J.~H.,  {Strohmayer} T.~E.,   {Zhang} W.,  2006, ApJS, 163, 401

\bibitem[\protect\citeauthoryear{{Jaisawal}, {Naik}  \& {Epili}}{{Jaisawal}
  et~al.}{2016}]{jaisawal2016a}
{Jaisawal} G.~K.,  {Naik} S.,   {Epili} P.,  2016, MNRAS, 457, 2749

\bibitem[\protect\citeauthoryear{{Kalberla}, {Burton}, {Hartmann}, {Arnal},
  {Bajaja}, {Morras}  \& {P{\"o}ppel}}{{Kalberla} et~al.}{2005}]{kalberla2005a}
{Kalberla} P.~M.~W.,  {Burton} W.~B.,  {Hartmann} D.,  {Arnal} E.~M.,  {Bajaja}
  E.,  {Morras} R.,   {P{\"o}ppel} W.~G.~L.,  2005, A\&A, 440, 775

\bibitem[\protect\citeauthoryear{{Klochkov} et~al.,}{{Klochkov}
  et~al.}{2012}]{klochkov2012a}
{Klochkov} D.,  et~al., 2012, A\&A, 542, L28

\bibitem[\protect\citeauthoryear{{Kopal}}{{Kopal}}{1959}]{kopal1959a}
{Kopal} Z.,  1959, Close binary systems.
The International Astrophysics Series, John Wiley \& Sons Inc., New York

\bibitem[\protect\citeauthoryear{{Larwood}}{{Larwood}}{1998}]{larwood1998a}
{Larwood} J.,  1998, MNRAS, 299, L32

\bibitem[\protect\citeauthoryear{{Lee}, {Osaki}  \& {Saio}}{{Lee}
  et~al.}{1991}]{lee1991a}
{Lee} U.,  {Osaki} Y.,   {Saio} H.,  1991, MNRAS, 250, 432

\bibitem[\protect\citeauthoryear{{Lewin}, {Clark}  \& {Smith}}{{Lewin}
  et~al.}{1968}]{lewin1968a}
{Lewin} W.~H.~G.,  {Clark} G.~W.,   {Smith} W.~B.,  1968, Nature, 219, 1235

\bibitem[\protect\citeauthoryear{{Liu}, {van Paradijs}  \& {van den
  Heuvel}}{{Liu} et~al.}{2006}]{liu2006a}
{Liu} Q.~Z.,  {van Paradijs} J.,   {van den Heuvel} E.~P.~J.,  2006, A\&A, 455,
  1165

\bibitem[\protect\citeauthoryear{{Lubow}, {Martin}  \& {Nixon}}{{Lubow}
  et~al.}{2015}]{lubow2015a}
{Lubow} S.~H.,  {Martin} R.~G.,   {Nixon} C.,  2015, ApJ, 800, 96

\bibitem[\protect\citeauthoryear{{Manousakis} et~al.,}{{Manousakis}
  et~al.}{2008}]{manousakis2008a}
{Manousakis} A.,  et~al., 2008, INTEGRAL hard X-ray detection of HMXB GX 304-1
  and H 1417-624, ATel 1613

\bibitem[\protect\citeauthoryear{{Martin}, {Pringle}, {Tout}  \&
  {Lubow}}{{Martin} et~al.}{2011}]{martin2011a}
{Martin} R.~G.,  {Pringle} J.~E.,  {Tout} C.~A.,   {Lubow} S.~H.,  2011, MNRAS,
  416, 2827

\bibitem[\protect\citeauthoryear{{Mason}, {Murdin}, {Parkes}  \&
  {Visvanathan}}{{Mason} et~al.}{1978}]{mason1978a}
{Mason} K.~O.,  {Murdin} P.~G.,  {Parkes} G.~E.,   {Visvanathan} N.,  1978,
  MNRAS, 184, 45P

\bibitem[\protect\citeauthoryear{{McClintock}, {Nugent}, {Li}  \&
  {Rappaport}}{{McClintock} et~al.}{1977}]{mcclintock1977a}
{McClintock} J.~E.,  {Nugent} J.~J.,  {Li} F.~K.,   {Rappaport} S.~A.,  1977,
  ApJ, 216, L15

\bibitem[\protect\citeauthoryear{{Miranda} \& {Lai}}{{Miranda} \&
  {Lai}}{2015}]{miranda2015b}
{Miranda} R.,  {Lai} D.,  2015, MNRAS, 452, 2396

\bibitem[\protect\citeauthoryear{{Moritani} et~al.,}{{Moritani}
  et~al.}{2013}]{moritani2013a}
{Moritani} Y.,  et~al., 2013, PASJ, 65, 83

\bibitem[\protect\citeauthoryear{{M\"uller}}{{M\"uller}}{2013}]{mueller2013a}
{M\"uller} S.,  2013, PhD thesis, Friedrich-Alexander-Universit\"at
  Erlangen-N\"urnberg

\bibitem[\protect\citeauthoryear{{Naoz}}{{Naoz}}{2016}]{naoz2016a}
{Naoz} S.,  2016, ARA\&A, 54, 441

\bibitem[\protect\citeauthoryear{{Naoz}, {Farr}, {Lithwick}, {Rasio}  \&
  {Teyssandier}}{{Naoz} et~al.}{2013}]{naoz2013a}
{Naoz} S.,  {Farr} W.~M.,  {Lithwick} Y.,  {Rasio} F.~A.,   {Teyssandier} J.,
  2013, MNRAS, 431, 2155

\bibitem[\protect\citeauthoryear{{Negueruela} \& {Okazaki}}{{Negueruela} \&
  {Okazaki}}{2001}]{negueruela2001a}
{Negueruela} I.,  {Okazaki} A.~T.,  2001, A\&A, 369, 108

\bibitem[\protect\citeauthoryear{{Nyman}}{{Nyman}}{2008}]{nyman2008a}
{Nyman} L.-{\AA}.,  2008, in {Reipurth} B.,  ed., , Handbook of Star Forming
  Regions, Volume II.
Astronomical Society of the Pacific Monograph Publications, p.~222

\bibitem[\protect\citeauthoryear{{Okazaki} \& {Negueruela}}{{Okazaki} \&
  {Negueruela}}{2001}]{okazaki2001a}
{Okazaki} A.~T.,  {Negueruela} I.,  2001, A\&A, 377, 161

\bibitem[\protect\citeauthoryear{{Okazaki}, {Bate}, {Ogilvie}  \&
  {Pringle}}{{Okazaki} et~al.}{2002}]{okazaki2002a}
{Okazaki} A.~T.,  {Bate} M.~R.,  {Ogilvie} G.~I.,   {Pringle} J.~E.,  2002,
  MNRAS, 337, 967

\bibitem[\protect\citeauthoryear{{Okazaki}, {Hayasaki}  \&
  {Moritani}}{{Okazaki} et~al.}{2013}]{okazaki2013a}
{Okazaki} A.~T.,  {Hayasaki} K.,   {Moritani} Y.,  2013, PASJ, 65, 41

\bibitem[\protect\citeauthoryear{{Papaloizou} \& {Lin}}{{Papaloizou} \&
  {Lin}}{1994}]{papaloizou1994a}
{Papaloizou} J.~C.~B.,  {Lin} D.~N.~C.,  1994, in {Duschl} W.~J.,  {Frank} J.,
  {Meyer} F.,  {Meyer-Hofmeister} E.,   {Tscharnuter} W.~M.,  eds,  NATO ASI
  Series Vol. 417, Theory of Accretion Disks - 2. Kluwer Academic Publishers,
  Dordrecht, Netherlands, p.~329

\bibitem[\protect\citeauthoryear{{Papaloizou} \& {Terquem}}{{Papaloizou} \&
  {Terquem}}{1995}]{papaloizou1995a}
{Papaloizou} J.~C.~B.,  {Terquem} C.,  1995, MNRAS, 274, 987

\bibitem[\protect\citeauthoryear{{Parkes}, {Murdin}  \& {Mason}}{{Parkes}
  et~al.}{1980}]{parkes1980a}
{Parkes} G.~E.,  {Murdin} P.~G.,   {Mason} K.~O.,  1980, MNRAS, 190, 537

\bibitem[\protect\citeauthoryear{{Paxton}}{{Paxton}}{2004}]{paxton2004a}
{Paxton} B.,  2004, PASP, 116, 699

\bibitem[\protect\citeauthoryear{{Postnov}, {Mironov}, {Lutovinov}, {Shakura},
  {Kochetkova}  \& {Tsygankov}}{{Postnov} et~al.}{2015}]{postnov2015a}
{Postnov} K.~A.,  {Mironov} A.~I.,  {Lutovinov} A.~A.,  {Shakura} N.~I.,
  {Kochetkova} A.~Y.,   {Tsygankov} S.~S.,  2015, MNRAS, 446, 1013

\bibitem[\protect\citeauthoryear{{Pottschmidt}, {Rothschild}, {Gasaway},
  {Suchy}  \& {Coburn}}{{Pottschmidt} et~al.}{2006}]{pottschmidt2006a}
{Pottschmidt} K.,  {Rothschild} R.~E.,  {Gasaway} T.,  {Suchy} S.,   {Coburn}
  W.,  2006, BAAS, 38, 384

\bibitem[\protect\citeauthoryear{{Priedhorsky} \& {Terrell}}{{Priedhorsky} \&
  {Terrell}}{1983}]{priedhorsky1983a}
{Priedhorsky} W.~C.,  {Terrell} J.,  1983, ApJ, 273, 709

\bibitem[\protect\citeauthoryear{{Rappaport}, {Clark}, {Cominsky}, {Li}  \&
  {Joss}}{{Rappaport} et~al.}{1978}]{rappaport1978a}
{Rappaport} S.,  {Clark} G.~W.,  {Cominsky} L.,  {Li} F.,   {Joss} P.~C.,
  1978, ApJ, 224, L1

\bibitem[\protect\citeauthoryear{{Reig}, {Fabregat}  \& {Coe}}{{Reig}
  et~al.}{1997}]{reig1997a}
{Reig} P.,  {Fabregat} J.,   {Coe} M.~J.,  1997, A\&A, 322, 193

\bibitem[\protect\citeauthoryear{{Reig}, {Larionov}, {Negueruela}, {Arkharov}
  \& {Kudryavtseva}}{{Reig} et~al.}{2007}]{reig2007a}
{Reig} P.,  {Larionov} V.,  {Negueruela} I.,  {Arkharov} A.~A.,
  {Kudryavtseva} N.~A.,  2007, A\&A, 462, 1081

\bibitem[\protect\citeauthoryear{{Rivinius}, {Carciofi}  \&
  {Martayan}}{{Rivinius} et~al.}{2013}]{rivinius2013a}
{Rivinius} T.,  {Carciofi} A.~C.,   {Martayan} C.,  2013, A\&ARv, 21, 69

\bibitem[\protect\citeauthoryear{{Rothschild} et~al.,}{{Rothschild}
  et~al.}{1998}]{rothschild1998a}
{Rothschild} R.~E.,  et~al., 1998, ApJ, 496, 538

\bibitem[\protect\citeauthoryear{{Rothschild} et~al.,}{{Rothschild}
  et~al.}{2017}]{rothschild2017a}
{Rothschild} R.~E.,  et~al., 2017, MNRAS, 466, 2752

\bibitem[\protect\citeauthoryear{{Sugizaki}, {Yamamoto}, {Mihara}, {Nakajima}
  \& {Makishima}}{{Sugizaki} et~al.}{2015}]{sugizaki2015b}
{Sugizaki} M.,  {Yamamoto} T.,  {Mihara} T.,  {Nakajima} M.,   {Makishima} K.,
  2015, PASJ, 67, 73

\bibitem[\protect\citeauthoryear{{Torres}, {Andersen}  \&
  {Gim{\'e}nez}}{{Torres} et~al.}{2010}]{torres2010a}
{Torres} G.,  {Andersen} J.,   {Gim{\'e}nez} A.,  2010, A\&ARv, 18, 67

\bibitem[\protect\citeauthoryear{{Verner}, {Ferland}, {Korista}  \&
  {Yakovlev}}{{Verner} et~al.}{1996}]{verner1996b}
{Verner} D.~A.,  {Ferland} G.~J.,  {Korista} K.~T.,   {Yakovlev} D.~G.,  1996,
  ApJ, 465, 487

\bibitem[\protect\citeauthoryear{{Walter}}{{Walter}}{1975}]{walter1975a}
{Walter} K.,  1975, A\&A, 42, 135

\bibitem[\protect\citeauthoryear{{Wilms}, {Allen}  \& {McCray}}{{Wilms}
  et~al.}{2000}]{wilms2000a}
{Wilms} J.,  {Allen} A.,   {McCray} R.,  2000, ApJ, 542, 914

\bibitem[\protect\citeauthoryear{{Wisniewski}, {Draper}, {Bjorkman}, {Meade},
  {Bjorkman}  \& {Kowalski}}{{Wisniewski} et~al.}{2010}]{wisniewski2010a}
{Wisniewski} J.~P.,  {Draper} Z.~H.,  {Bjorkman} K.~S.,  {Meade} M.~R.,
  {Bjorkman} J.~E.,   {Kowalski} A.~F.,  2010, ApJ, 709, 1306

\bibitem[\protect\citeauthoryear{{Yamamoto}, {Sugizaki}, {Mihara}, {Nakajima},
  {Yamaoka}, {Matsuoka}, {Morii}  \& {Makishima}}{{Yamamoto}
  et~al.}{2011}]{yamamoto2011a}
{Yamamoto} T.,  {Sugizaki} M.,  {Mihara} T.,  {Nakajima} M.,  {Yamaoka} K.,
  {Matsuoka} M.,  {Morii} M.,   {Makishima} K.,  2011, PASJ, 63, S751

\bibitem[\protect\citeauthoryear{{Zamanov}, {Reig}, {Mart{\'{\i}}}, {Coe},
  {Fabregat}, {Tomov}  \& {Valchev}}{{Zamanov} et~al.}{2001}]{zamanov2001a}
{Zamanov} R.~K.,  {Reig} P.,  {Mart{\'{\i}}} J.,  {Coe} M.~J.,  {Fabregat} J.,
  {Tomov} N.~A.,   {Valchev} T.,  2001, A\&A, 367, 884

\makeatother
\end{thebibliography}

\bsp

\appendix
\section{Stellar precession in binaries}\label{app:precess}

\begin{figure}
  \includegraphics[width=\columnwidth]{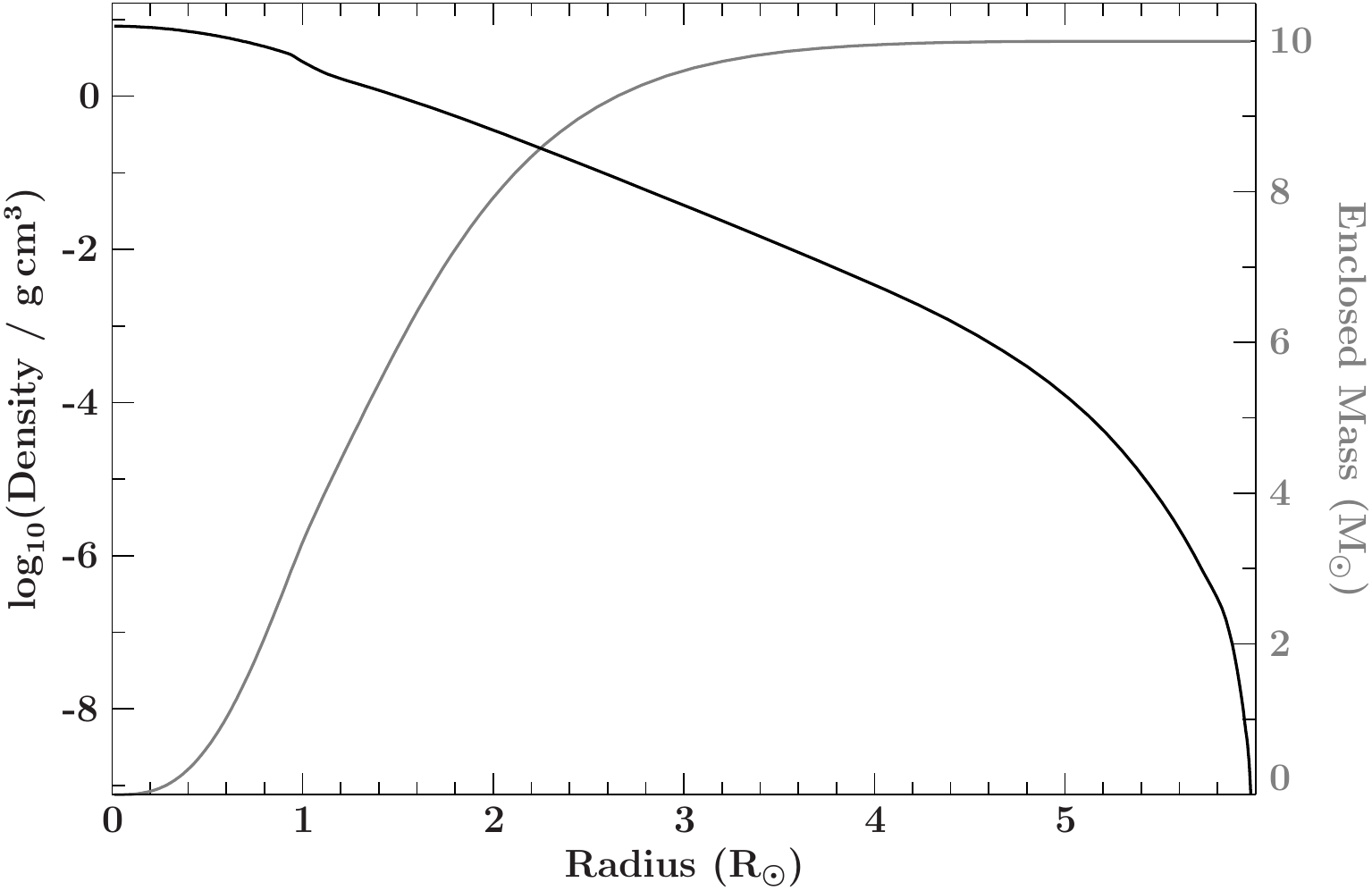}
  \caption{Used density profile (black) and enclosed mass (gray) of a
    star as a function of its stellar radius as calculated with the EZ
    code.}
  \label{fig:bestardensity}
\end{figure}

In BeXRBs, if the rotation axis of the Be star is inclined with
respect to the orbital plane, the compact object will exert a tidal
torque on the Be star and disk. The torque on the Be star will be
non-zero if the Be star is oblate (due to, e.g., fast rotation). In
the following we estimate the resulting precession period of the Be
star using two formulations by \citet[hereafter K59]{kopal1959a} and
by \citet[hereafter A76]{alexander1976a}. For all calculations we
assume the orbital parameters as listed in Table~\ref{tab:orbit}, a
stellar mass and radius for the Be star of $M_* = 10\,\msol$ and
$R_* = 6\,\rsol$, respectively (taken from Table~\ref{tab:params}),
and a neutron star mass of $M_{X} = 1.4\,\msol$. The radial density
profile of the Be star as shown in Fig.~\ref{fig:bestardensity} was
calculated using the Evolve ZAMS (EZ) code\footnote{We have used the
  online tool EZ-Web developed by Richard Townsend,
  \url{http://www.astro.wisc.edu/~townsend/static.php?ref=ez-web}.}
\citep{paxton2004a} for an initial stellar mass and metallicity of
$10\,\msol$ and 0.02, respectively. We simulated stellar evolution up
to an age of 15.6\,Myr in order to let the star expand up to the
assumed radius of $6\,\rsol$. When discussing other stellar radii or
masses as shown in Fig.~\ref{fig:k59a76}, we have linearly scaled the
density profile for simplicity. In case of a different stellar mass,
we scaled the density such that the total enclosed mass, i.e., the
integral of the density profile over the full radius results in the
assumed mass.

\subsection{Formulation by \citetalias{kopal1959a}}
\label{app:precess:K59}

\begin{figure*}
  \includegraphics[width=2\columnwidth]{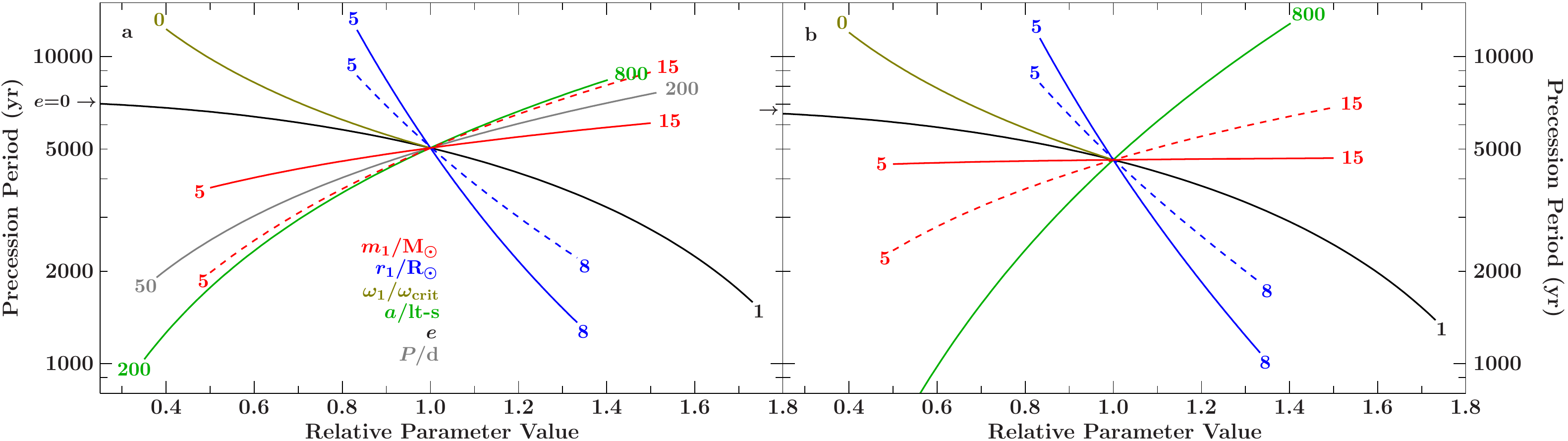}
  \caption{Dependencies of the precession period on stellar and binary
  parameters after \textbf{a)} \citetalias{kopal1959a} and \textbf{b)}
  \citetalias{alexander1976a}. The parameter values are normalized to
  the ones assumed for \gx as given in the text. The parameter ranges
  are provided by absolute values at the line endings. The dashed lines
  have been calculated taking a linearly scaled density profile into
  account (see Fig.~\ref{fig:bestardensity} and the text for details).
  The arrow on the y-axis labelled $e{=}0$ marks the value of the
  precession period for a circular orbit, to which the corresponding
  line (black) converges.} \label{fig:k59a76}
\end{figure*}

Assuming that the inclination of the rotation axis of the primary star
to the binary orbit axis is small, \citetalias{kopal1959a} derived the
period of stellar precession as \citepalias[Eq.~II.5-62]{kopal1959a}
\begin{equation}
  \frac{P_\mathrm{orb}}{U_1} = \Gamma_1 + \Pi_1,
  \label{eq:5-62}
\end{equation}
where $P_\mathrm{orb}$ is the orbital period, $U_1$ is the period of
precession of the primary star, and $\Gamma_1$ and $\Pi_1$ are
quantities defined by \citepalias[Eq.~II.5-25]{kopal1959a}
\begin{equation}
  \Gamma_1 = \frac{3}{2}\frac{C_1^{\prime\prime}}{C_1^\prime} \gamma_1
  \label{eq:5-25}
\end{equation}
and \citepalias[Eq.~II.5-33]{kopal1959a}
\begin{equation}
  \Pi_1 = \frac{3}{4 a^2}\frac{C_1^\prime-A_1^\prime}{m_1}.
  \label{eq:5-33}
\end{equation}
Here, $m_1$ is the mass of the primary, $a$ is the semi-major axis of
the binary, and $\gamma_1$ is the ratio of the rotation frequency of
the primary, $\omega_1$, to the binary orbital frequency
\citepalias[Eq.~II.5-11]{kopal1959a},
\begin{equation}
  \gamma_1 = \frac{\omega_1}{n} = \omega_1 \sqrt{\frac{a^3}{G(m_1+m_2)}},
  \label{eq:5-11}
\end{equation}
with the mass, $m_2$, of the secondary and the gravitational constant,
$G$. The moment of inertia of the primary perpendicular, $A_1^\prime$,
and parallel, $C_1^\prime$, to its rotational axis are given by
\citepalias[Eq.~II.3-34]{kopal1959a}
\begin{equation}
  A_1^\prime = \frac{8}{3}\pi \int_0^{r_1} \rho {r^\prime}^4 \mathrm{d}r^\prime
  - \frac{(\Delta_2-1) \omega_1^2 r_1^5}{9G}
  \label{eq:3-34}
\end{equation}
and by \citepalias[Eq.~II.3-39]{kopal1959a}
\begin{equation}
  C_1^\prime = \frac{8}{3}\pi \int_0^{r_1} \rho {r^\prime}^4 \mathrm{d}r^\prime
  + \frac{2(\Delta_2-1) \omega_1^2 r_1^5}{9G},
  \label{eq:3-39}
\end{equation}
respectively. Here, $\rho$ is the density of the primary star as a
function of the distance, $r^\prime$, to its centre up to the full
radius, $r_1$. In each equation, the first term on the right hand side
is the moment of inertia of the primary at rest, which is calculated for
the stellar structure as shown in Fig.~\ref{fig:bestardensity}. The
second term describes the contribution of the rotational deformation to
the moment of inertia, where $\Delta_2$ is related to the apsidal
constant, $k_2$, as
\begin{equation}
  \Delta_2 = 1 + k_2 .
  \label{eq:Delta_2}
\end{equation}
We assume $k_2 = 0.01$ given that the apsidal constant of $10\,\msol$
main-sequence stars are typically in the range of (5--8)$\times
10^{-3}$ \citep{torres2010a}.

The tidal bulge forming inside the primary star due to the gravitational
pull of the secondary modifies the moment of inertia. In the rotating
reference frame of the orbital plane, this moment of inertia,
$C_1^{\prime\prime}$, is calculated by
\citepalias[Eq.~II.3-46]{kopal1959a}
\begin{equation}
  C_1^{\prime\prime} \simeq \frac{1}{3} (\Delta_2-1) \frac{m_2 r_1^5}{d^3},
  \label{eq:3-46}
\end{equation}
where $d$ is the distance of the secondary from the primary, which is
given by
\begin{equation}
  d = \frac{a (1-e^2)}{1+e \cos f},
  \label{eq:orbit}
\end{equation}
with $e$ and $f$ being the eccentricity and the true anomaly. When orbit
averaged, Eq.~\ref{eq:3-46} is reduced to
\begin{equation}
  C_1^{\prime\prime} \simeq \frac{(\Delta_2-1) m_2
	r_1^5}{3 a^3 (1-e^2)^{3/2}}.
  \label{eq:3-46b}
\end{equation}
Using Eqs.~\ref{eq:5-11}--\ref{eq:3-39}, the orbit average of
$C_1^{\prime\prime}$ can also be written as
\begin{equation}
  C_1^{\prime\prime} \simeq (C_1^\prime - A_1^\prime)
  \frac{G m_2}{\omega_1^2 a^3 (1-e^2)^{3/2}} = \frac{m_2}{m_1+m_2}
  \frac{(C_1^\prime - A_1^\prime)}{(1-e^2)^{3/2} \gamma_1^2}.
  \label{eq:3-46c}
\end{equation}
Note that the factor $(1-e^2)^{3/2}$ is set to 1 in
\citetalias{kopal1959a}, where the orbit is assumed to be almost
circular. We retain this factor in Eq.~\ref{eq:3-46c} since the orbits
of BeXRBs are generally eccentric.

With Eq.~\ref{eq:3-46c}, we can write Eq.~\ref{eq:5-25} as
\begin{equation}
  \Gamma_1 = \frac{3}{2}\frac{C_1^{\prime\prime}}{C_1^\prime} \gamma_1 =
  \frac{3}{2}\frac{m_2}{m_1+m_2} \frac{C_1^\prime -
  A_1^\prime}{C_1^\prime} (1-e^2)^{-3/2} \gamma_1^{-1}. \label{eq:5-25b}
\end{equation}
Note that the power of $\gamma_1$ in Eq.~\ref{eq:5-25b} of $-1$ is
different than in the original Eq.~II.5-25 in \citetalias{kopal1959a}
of $+1$, which has been noticed by, e.g., \citet{walter1975a} already.

Assuming that the Be star is rotating at the critical speed,
$\omega_\mathrm{crit}$, as an upper
limit,
\begin{equation}
  \omega_1 = \omega_\mathrm{crit} = \sqrt{G m_1/r_1^3},
\end{equation}
and setting $r_1=R_*$, $m_1=M_*$, and $m_2=M_\mathrm{x}$ we finally find
$U_1 \sim 5000$\,yr for the precession period of the Be star in \gx as
calculated after Eq.~\ref{eq:5-62}. The dependencies of
Eq.~\ref{eq:5-62} on the stellar and binary parameters is shown in
Fig.~\ref{fig:k59a76}a.

\subsection{Formulation by \citetalias{alexander1976a}}
\label{app:precess:A76}

\citet{alexander1976a} generalized the formulation of \citet{kopal1959a}
to an arbitrary angle, $\Theta$, between the angular momentum vectors of
the primary and the binary orbit. If the angular momentum vectors
of the Be star and its disk align, $\Theta$ is then equal to the disk
misalignment angle, $\delta$, introduced in
Sect.~\ref{sec:densitymodel}. In comparison to Eq.~\ref{eq:5-62},
\citet{alexander1976a} derived \citepalias[Eq.~2.21]{alexander1976a}
\begin{equation}
  -\mu_1 h_1 \cos \Theta \frac{H}{H_0} = \frac{2 \pi}{U_1},
\label{eq:2.21}
\end{equation}
for the precession period, $U_1$, of the primary, which is negative for
retrograde precession. Here, $\mu_1$ is a constant given by
\citepalias[Eq.~2.18]{alexander1976a}
\begin{equation}
  \mu_1 = \frac{1}{C_1 + 2C_\mathrm{R1}-C_\mathrm{T1}}
  \frac{3C_\mathrm{T1}}{C_1+3C_\mathrm{T1}}, \label{eq:2.18}
\end{equation}
with the moment of inertia, $C_1$, of the primary star at rest, which is
calculated by \citepalias[Eq.~2.4]{alexander1976a}
\begin{equation}
  C_1 = \frac{8}{3}\pi \int_0^{r_1} \rho {r^\prime}^4 \mathrm{d}r^\prime
  \label{eq:2.4}
\end{equation}
and equal to the first term on the right hand side in Eq.~\ref{eq:3-39}.
The rotational and tidal distortions to the total moment of inertia are
are given by \citepalias[Eq.~2.5]{alexander1976a}
\begin{equation}
  C_\mathrm{R1} = \frac{k_2 \omega_1^2 r_1^5}{9G}
  \label{eq:2.5}
\end{equation}
and \citepalias[Eq.~2.14]{alexander1976a}
\begin{equation}
  C_\mathrm{T1} = \frac{1}{6} k_2 \frac{m_2 r_1^5}{a^3(1-e^2)^{3/2}},
  \label{eq:2.14}
\end{equation}
respectively. 

The total angular momentum, $H$, of the binary is defined as
\citepalias[Eq.~2.19]{alexander1976a}
\begin{equation}
  \vec{h}_1 + \vec{h}_2 + H_0 \vec{s}_0 = \vec{H}.
  \label{eq:2.19}
\end{equation}
The angular momentum of the primary, $\vec{h}_1$, is expressed
as \citepalias[Eq.~2.12]{alexander1976a}
\begin{equation}
  \vec{h}_1 = (C_1 + 2C_\mathrm{R1}-C_\mathrm{T1}) \vec{\omega}_1 +
  3C_\mathrm{T1} (\vec{\omega}_1 \cdot \vec{s}_0) \vec{s}_0,
  \label{eq:2.12}
\end{equation}
with its angular velocity vector, $\vec{\omega}_1$, and the normal
vector, $\vec{s}_0$, of the orbital plane. Since the secondary is a
compact object, which is a point mass in good approximation, its angular
momentum is $\vec{h}_2 = \vec{0}$. The orbital angular momentum of the
binary, $H_0$, is given by \citepalias[Eq.~2.11]{alexander1976a}
\begin{equation}
  \vec{H}_0 = H_0 \vec{s}_0 = \frac{m_1 m_2}{m_1 +
  m_2}\sqrt{G(m_1+m_2)a(1-e^2)} \vec{s}_0, \label{eq:2.11}
\end{equation}

Assuming a misalignment angle $\Theta = 0$ as an upper limit and the
same parameters for \gx as previously in Sect.~\ref{app:precess:K59}, we
find a precession period of $U_1 \sim - 4600$\,yr using
Eq.~\ref{eq:2.21}, which agrees with the estimation following
\citetalias{kopal1959a}. Fig.~\ref{fig:k59a76}b shows the dependencies
of Eq.~\ref{eq:2.21} on the stellar and binary parameters.

\subsection{Parameter dependencies}

The precession period of the stellar companion star in \gx, which we
derived following \citetalias{kopal1959a} and
\citetalias{alexander1976a}, is around 5000\,yr and both formulations
agree within a few hundred years. This period does not explain the short
period of ${\sim}2$\,yr we have discovered by modelling the evolution of
the absorption column density (see Sect.~\ref{sec:model}), which we
interpret as the precession period of the Be disk.

Since we have fixed the stellar and binary parameters for the estimation
of the stellar precession period to the values given above, different
parameter values might change the resulting period significantly.
Figure~\ref{fig:k59a76} shows the dependency of the stellar precession
period after \citetalias{kopal1959a} and \citetalias{alexander1976a} on
these parameters within a reasonable range. As can be seen, the
precession period is larger than 1000\,yr in almost all cases. Thus, it
is unlikely that a certain combination of parameters indeed results in
a precession period consistent with our findings.

We note that there is a difference between the formulations by
\citetalias{kopal1959a} and \citetalias{alexander1976a}, which is most
prominent in the dependency on the primary mass, $m_1$ (red curve in
Fig.~\ref{fig:k59a76}), and the semi-major axis, $a$ (green curve). This
is due to the fact that \citetalias{alexander1976a} uses Kepler's third
law,
\begin{equation}
  \frac{P_\mathrm{orb}^2}{a^3} = \frac{4 \pi^2}{G (m_1 +
  m_2)},\label{eq:kepler3}
\end{equation}
to calculate the binary orbital period, $P_\mathrm{orb}$, while the
equations in \citetalias{kopal1959a} have a direct dependency. Indeed,
when substituting $P_\mathrm{orb}$ in Eq.~\ref{eq:5-62} with
Eq.~\ref{eq:kepler3}, both formulations agree very well.


\end{document}